\documentclass[journal]{IEEEtran} 

\usepackage{verbatim}
\usepackage{epsfig,scalefnt,multirow}
\usepackage{url}
\usepackage{ifthen}
\usepackage{mathtools}
\usepackage{cite}
\usepackage{graphicx}
\usepackage{amssymb}
\usepackage{caption}
\usepackage{booktabs}
\usepackage{amsmath}
\usepackage{epstopdf}
\usepackage{algorithm}
\usepackage{algpseudocode}
\usepackage{cases}
\usepackage{xcolor}
\usepackage{multicol}
\usepackage{stfloats}
\usepackage{blindtext}
\usepackage{caption}
\usepackage{subcaption}

\ifCLASSINFOpdf
\else
\fi
   \def\cH{{\mathcal{H}}}
\def\cI{{\mathcal{I}}}

\def\argmax{\mathop{\mathrm{argmax}}}

\def\Re{\mathop{\mathrm{Re}}}
\def\Im{\mathop{\mathrm{Im}}}

\def\b0{{\pmb{0}}} 

   \def\bd{{\mathbf{d}}}
\def\bee{{\mathbf{e}}}   
   
   \def\bp{{\mathbf{p}}}

\def\bA{{\mathbf{A}}}   
   \def\bH{{\mathbf{H}}}

   \def\bX{{\mathbf{X}}}

\DeclarePairedDelimiter\norm{\lVert}{\rVert}

\def\blue{\textcolor{blue}}

\IEEEoverridecommandlockouts
\begin{document}
%
\title{Meta-Learning-Based People Counting and Localization Models Employing CSI from Commodity WiFi NICs
\thanks{This work was supported in part by the Ministry of Science and ICT (MSIT), Korea, under the Information Technology Research Center (ITRC) support program (IITP-2025-2020-0-01787) supervised by the Institute of Information \& Communications Technology Planning \& Evaluation (IITP) and by Institute of Information \& Communications Technology Planning \& Evaluation (IITP) under 6G·Cloud Research and Education Open Hub (IITP-2025-RS-2024-00428780) grant funded by the Korea government (MSIT). (\textit{Jihoon Cha and Hwanjin Kim contributed equally to this work.})}
\thanks{J. Cha and J. Choi are with the School of Electrical Engineering, Korea Advanced Institute of Science and Technology, Daejeon 34141, South Korea (e-mail: charge@kaist.ac.kr; junil@kaist.ac.kr).
	
Hwanjin Kim is with the School of Electronics Engineering, Kyungpook National University, Daegu 41566, South Korea (e-mail:
hwanjin@knu.ac.kr).}}
%
%
%

\author{Jihoon Cha, \IEEEmembership{Graduate Student Member,~IEEE}, Hwanjin Kim, \IEEEmembership{Member,~IEEE}, \\ and Junil Choi, \IEEEmembership{Senior Member,~IEEE}}

\maketitle

\begin{abstract}
In this paper, we consider people counting and localization systems exploiting channel state information (CSI) measured from commodity WiFi network interface cards (NICs). While CSI has useful information of amplitude and phase to describe signal propagation in a measurement environment of interest, CSI measurement suffers from offsets due to various uncertainties. Moreover, an uncontrollable external environment where other WiFi devices communicate each other induces interfering signals, resulting in erroneous CSI captured at a receiver. In this paper, preprocessing of CSI is first proposed for offset removal, and it guarantees low-latency operation without any filtering process. Afterwards, we design people counting and localization models based on pre-training. To be adaptive to different measurement environments, meta-learning-based people counting and localization models are also proposed. Numerical results show that the proposed meta-learning-based people counting and localization models can achieve high sensing accuracy, compared to other learning schemes that follow simple training and test procedures.
\end{abstract}


%
\IEEEpeerreviewmaketitle

\section{Introduction}\label{sec1}
\IEEEPARstart{P}ROMISING techniques such as smart home and Internet of things are gradually realized in our daily lives, as electronic devices and even furniture can access to a unified network with wireless communications. For such devices to cooperatively provide personalized services for their surrounding users, wireless human sensing in indoor environments becomes critical \cite{Teixeira:2010,Liu:2019,Liu:2020a,Liu:2020b}. Based on the idea that the signals affected by objects and people record on-site information, human sensing has been studied by using various methods. Camera-based methods employ cameras to recognize human activities \cite{Yang:2003,Aggarwal:2011,Ke:2013}. They do not operate to dark and blind spots and have to handle privacy issues though. Radar-based systems take advantage of wide bandwidth to guarantee high resolution of sensing, and they can extract useful information even in a blind manner \cite{Adib:2015,Lien:2016,Alizadeh:2019}. However, deploying specialized hardware is not preferred as a cost- and power-effective solution in a daily situation. On the contrary, WiFi-based sensing exploits WiFi access points (APs), which are widely deployed in advance, without any additional hardware deployment.

Human sensing with WiFi signals was studied for various purposes, e.g., occupancy and people counting \cite{Khan:2023,Soltanaghaei:2020,Zhao:2019}, gesture and activity recognition \cite{Wang:2017a,Jiang:2020,Yousefi:2017,Bu:2020,Niu:2021,Arshad:2019}, and localization \cite{Noelia:2021,Foliadis:2021,Qian:2017,Qian:2018}, and some researches considered them simultaneously \cite{Tan:2019,Choi:2021,Yan:2021}. All these tasks are based on WiFi channel measurements, for which commodity WiFi network interface cards (NICs) are usually employed. NICs can provide two kinds of channel information from the measurement, namely, received signal strength indicator (RSSI) and channel state information (CSI) \cite{Liu:2020a}. Both of them describe the environment with moving objects and static obstacles, but CSI provides larger dimensional information than RSSI. CSI contains phase information of channels for multiple subcarriers as well as their amplitudes. Since the amplitude and phase information is distorted by offsets, it is required to reduce the offsets to fully exploit CSI \cite{Xie:2019, Wang:2017c}.

Some studies investigated the relation between the signal propagation and the formation of CSI and then estimated the parameters such as angle of arrival, Doppler frequency shift, and time of flight \cite{Qian:2017,Qian:2018,Tan:2019,Niu:2021}. They were able to provide geometrical information of people serving as reflector or blockage, but such parameter-based analyses become infeasible in more complicated situations. Instead, as a high-performance computing system can manage to process a massive amount of data, deep neural network (DNN) is considered as another practical solution to handle sensing problems \cite{Yousefi:2017,Zhao:2019,Wang:2017b,Yan:2021}. With conducting only simple preprocessing, DNN can employ entire CSI as model inputs with minimizing potential loss of channel information by parameter extraction.

While DNN can be utilized to solve the human sensing problem in ordinary indoor environments, its performance is usually dependent of the domains where and when the training and test samples are collected. In other words, although a DNN model achieves very high accuracy for a certain place and time duration, it may not work in different environments.
Since DNN commonly uses a huge number of samples for training, repeated training for new environments is time-consuming. In this case, transfer learning can adapt the pre-trained model to a new environment with a small number of training samples \cite{Khan:2023,Noelia:2021,Soltanaghaei:2020,Bu:2020,Arshad:2019}.

In this paper, we propose preprocessing of CSI measured from commodity WiFi NICs. This process is to conduct offset removal with simple and low-latency operation and does not require any filtering. As baseline, we explain pre-training-based people counting and localization models with the preprocessed CSI. To achieve high adaptability to test environments different from training environments, we propose meta-learning-based people counting and localization models. The proposed model is based on the model-agnostic meta-learning (MAML) algorithm, which is universally employed by adapting the learned model to each of specialized tasks \cite{Finn:2017}. Compared to other DNN structures, the proposed model using only a small number of adaptation samples after pre-training can achieve high accuracy for different test environments. Through experiments in several environments, we validate the applicability of meta-learning-based model to solve the human sensing problem with CSI measured by commodity WiFi NICs.

The paper is organized as follows. In Section \ref{sec2}, we briefly describe a WiFi signal transmission model, explain a limitation of parameter-based approach, and introduce our human sensing process. Including a base module for learning models, pre-training-based people counting and localization models are introduced as preliminaries in Section \ref{sec3}. In Section \ref{sec4}, we elaborate on preprocessing of CSI that reduces CSI distortion to derive distinct features as model inputs for learning, then propose the meta-learning-based scheme for people counting and localization. In Section \ref{sec5}, we first analyze the computational complexity of the proposed meta-learning-based model, and then present the experimental setup and results to evaluate its performance in comparison to other DNN models. We finally make the conclusion in Section \ref{sec6}.

\textbf{Notations:} Lower and upper boldface letters represent respectively column vectors and matrices. $\bA^{\mathrm{T}}$ denotes the transpose of a matrix $ \bA $, and $\bA^{\dagger}$ is the pseudo-inverse matrix of $ \bA $. The $ (m,n) $-th element of $ \bA $ is denoted by $ [\bA]_{(m,n)} $. $ [\bA]_{(m,:)} $ and $ [\bA]_{(:,n)} $ represent the $ m $-th row and $ n $-th column of $ \bA $, respectively. $\mathbf{1}_N$ is used for the $N\times 1$ all one vector.  $ \Re\{z\} $ and $ \Im\{z\} $ represent the real and imaginary parts of a complex value $ z \in \mathbb{C}$ where $ \mathbb{C} $ is the set of all complex~numbers. $ \lvert z\rvert $ and $ \angle (z) $ are the absolute value and angle of $ z\in \mathbb{C} $, respectively. 

\section{System Model}\label{sec2}
We first explain the signal transmission model with WiFi NICs and erroneous CSI induced by uncontrollable factors. After commenting on the limitation to employ a conventional parameter-based approach for the erroneous CSI, we present overall process of proposed people counting and localization models.

\subsection{WiFi Signal Transmission Model}\label{sec2-1}
\begin{figure}
	\centering
	\includegraphics[width=0.95\columnwidth]{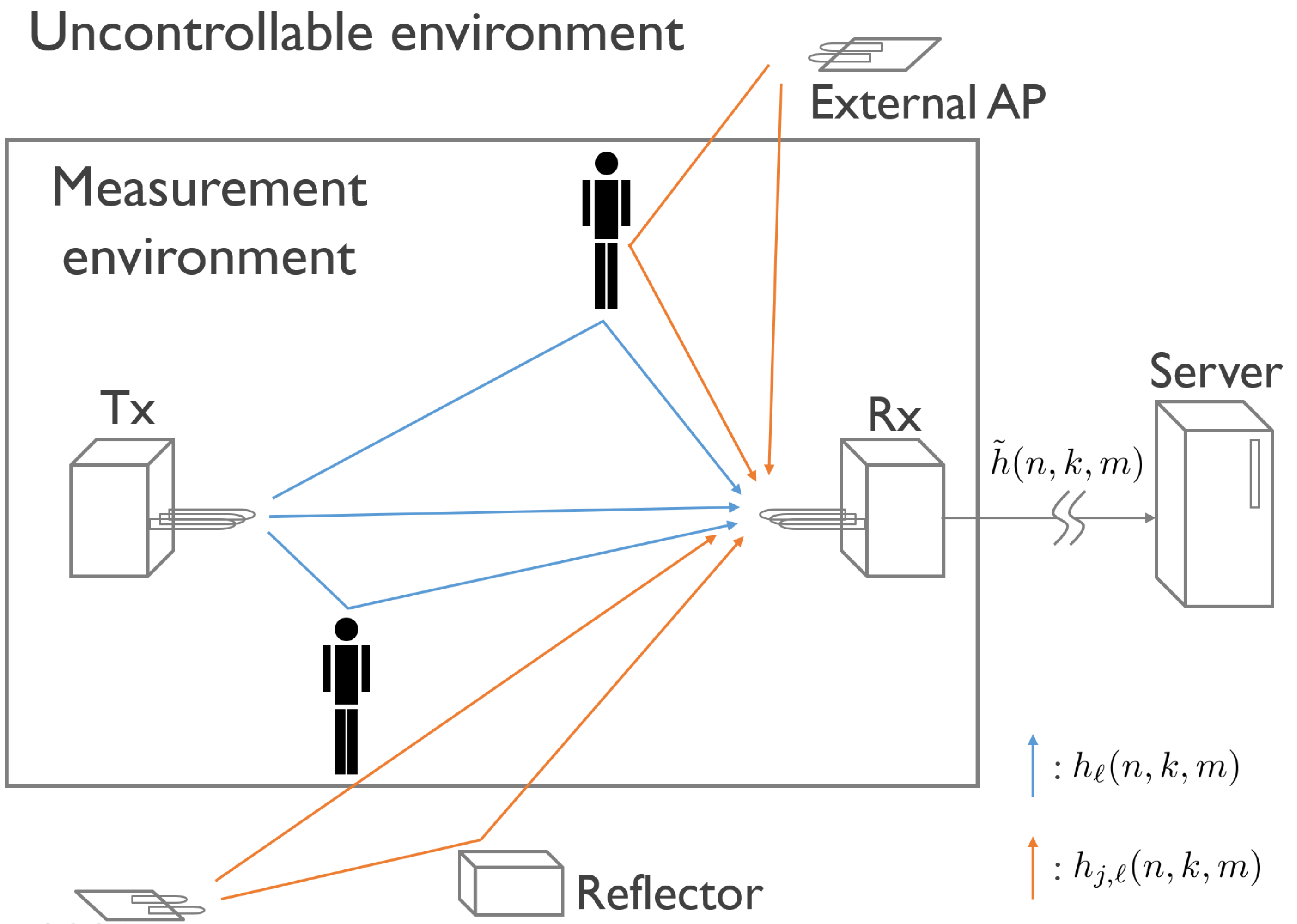}
	\caption{Structure of a CSI measurement system using commodity WiFi NICs for people counting and localization.} \label{fig1}
\end{figure}

Two commodity WiFi NICs connected to personal computers (PCs) serve as a transmitter and a receiver as in Fig. \ref{fig1}. The receiver with $ M_\mathrm{r} $ antennas measures CSI for $ K $ subcarriers based on each packet sent from the transmitter with $ M_\mathrm{t} $ antennas. The PC equipped with the receiver sends the measured CSI to a server, which conducts people counting and localization exploiting preprocessed~CSI.

The WiFi CSI of interest for the $ n $-th packet, the $ k $-th subcarrier, and the $ m $-th spatial link for $ {m \in \{1,2,\cdots,M_\mathrm{r}M_\mathrm{t}(\triangleq M)\}}$ can be modeled as
\begin{align}\label{channel}
h(n,k,m) &= \sum_{\ell=0}^{L_0-1}h_\ell(n,k,m)\notag\\
&=\sum_{\ell=0}^{L-1} \alpha_\ell(n,k,m) e^{-j2\pi f_k \tau_\ell(n,k,m)},
\end{align}
where\footnote{While the number of propagation paths can be different with $n$, $k$, and $m$ in practice, we assume it is the same for all paths to simplify the notation.} $ L_0 $ is the number of propagation paths by reflectors, $ h_\ell(n,k,m) $ is the channel corresponding to the $ \ell $-th path, and $ \alpha_\ell(n,k,m) $ represents the channel attenuation. The $ k $-th subcarrier frequency $ f_k $ is expressed~as
\begin{align}\label{subcarrier frequency}
f_k=f_c+\Delta f_k,
\end{align}
where $ f_c $ is the carrier frequency, and $ \Delta f_k $ is the frequency difference between $ f_k $ and $ f_c $. The propagation delay for the $ \ell $-th path $ \tau_\ell (n,k,m) $ is defined~as
\begin{align}\label{propagation delay}
\tau_\ell(n,k,m)=\bar{\tau}_\ell+\frac{f_{D,\ell}}{f_k}\Delta t_n+\frac{\Delta d_{m_\mathrm{r}} \sin\theta_\ell + \Delta d_{m_\mathrm{t}} \sin\phi_\ell}{v_\mathrm{c}},
\end{align}
where $ \bar{\tau}_\ell $ is the reference delay for the $ \ell $-th path, and $ f_{D,\ell} $ is the Doppler frequency shift by the $ \ell $-th moving reflector. The time duration for movement is denoted by $ \Delta t_n $, and $v_\mathrm{c}$ represents the speed of light. For the $ m $-th spatial link, the spatial difference between the $ m_\mathrm{r} $-th receive antenna and the reference receive antenna is represented by $ \Delta d_{m_\mathrm{r}} $, and $ \theta_\ell $ is the angle of arrival at the receiver. Similarly, the spatial difference and the angle of departure for the $ m_\mathrm{t} $-th transmit antenna are denoted by $ \Delta d_{m_\mathrm{t}} $ and $ \phi_\ell $, respectively.

There are two main factors inducing the CSI measured at the receiver to be erroneous: interference and offsets. For an indoor environment, there can be other WiFi APs and routers for general use. Assuming that they operate with the same or adjacent channel where the WiFi NICs communicate, co-channel interference is added to the CSI of interest. Reflectors outside the measurement environment also influence signal propagation arriving at the receiver. The interfering CSI by $ J $ other WiFi devices can be written as
\begin{align}\label{interference}
h_I(n,k,m) &= \sum_{j=1}^{J}\sum_{\ell=0}^{L_j-1}h_{j,\ell}(n,k,m)\notag\\
&=\sum_{j=1}^{J}\sum_{\ell=0}^{L_j-1} \alpha_{j,\ell}(n,k,m) e^{-j2\pi f_k \tau_{j,\ell}(n,k,m)},
\end{align}
where $ L_j $ is the number of propagation paths for each $ j\in\{1,2,\cdots,J\} $, and $ \alpha_{j,\ell}(n,k,m) $ and $ \tau_{j,\ell}(n,k,m) $ are the channel attenuation and the propagation delay for the interfering signals, respectively.

In addition to the co-channel interference, the CSI is distorted due to the offsets by imperfection of transceivers and their asynchronization. The erroneous CSI measured at the receiver can be modeled as \cite{Xie:2019, Wang:2017c, Speth:1999}
\begin{align}\label{erroneous CSI}
\tilde{h}(n,k,m) =&e^{j2\pi\left(\left(\epsilon_{b}+\epsilon_{s}(n)\right)k+\epsilon_c(n) \right)}\notag\\
&\times\left(h(n,k,m)+h_I(n,k,m)\right)+\epsilon_a(n,k),
\end{align}
where $ \epsilon_b $ and $ \epsilon_{s}(n) $ are the phase offsets from the packet boundary detection uncertainty and the sampling time difference of transceivers, respectively. The carrier frequency difference of transceivers causes the phase offset $ \epsilon_{c}(n) $, and $ \epsilon_a(n,k) $ is additional complex-valued noise. The two offsets $ \epsilon_b $ and $ \epsilon_{s}(n) $ induce subcarrier phase rotation proportional to $ k $, and the degree of rotation also changes steadily in time by $ \epsilon_{s}(n) $ and~$ \epsilon_{c}(n) $.

\subsection{General Sensing Process}
One direct approach for human sensing with CSI is to estimate the signal parameters indicating the existence and movement of the reflectors such as the propagation delay, the Doppler frequency shift, and the angle of arrival. However, existing methods based on this approach are only valid in a single-person environment. They are not proper for multiple-person scenarios, which are the focus in this paper, since it is almost infeasible to separately estimate the signal parameters for each person of interest. Moreover, CSI with co-channel interference and offsets makes it difficult to exploit parameter-based approaches without complicated preprocessing.

\begin{figure}
	\centering
	\includegraphics[width=0.95\columnwidth]{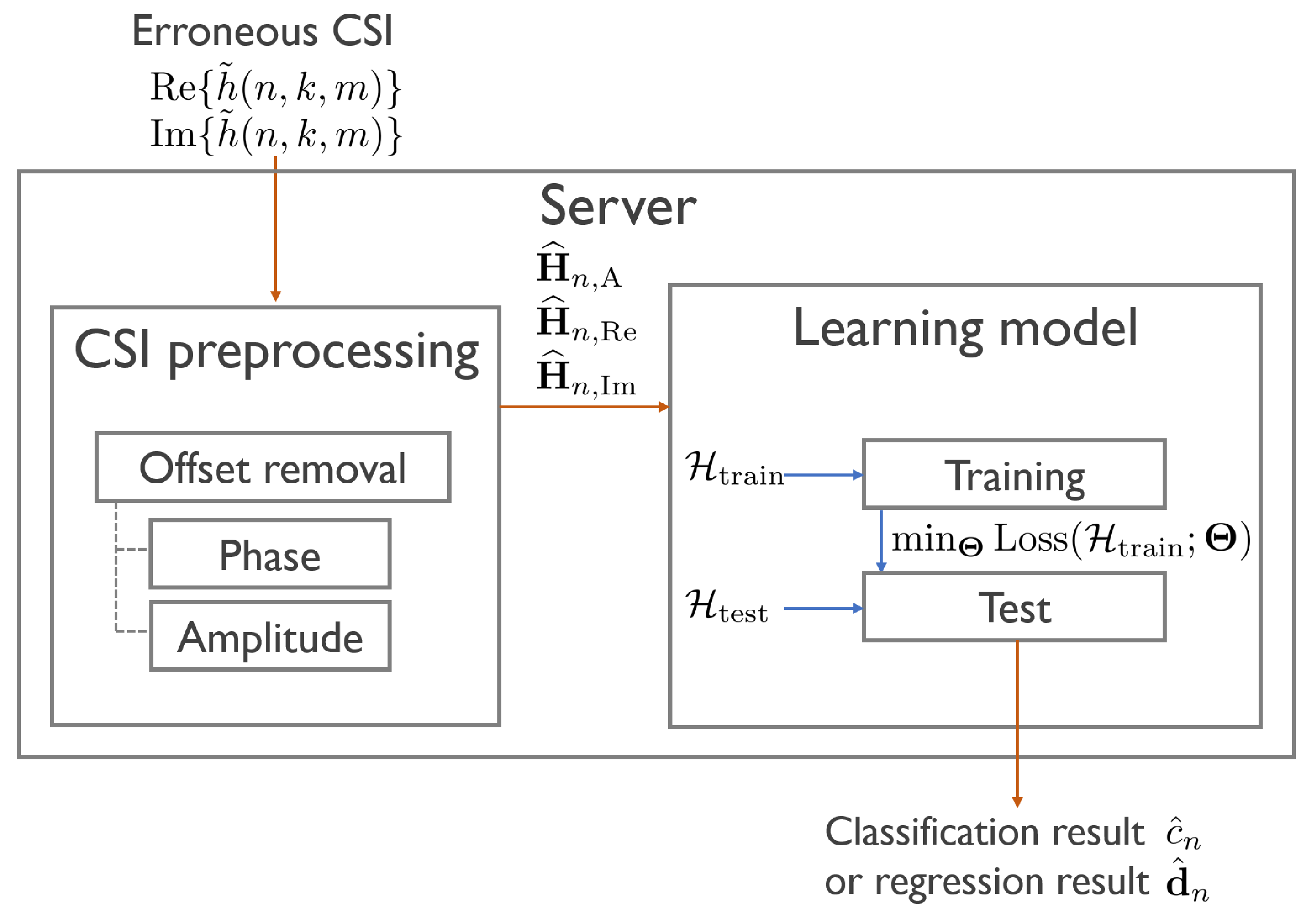}
	\caption{A remote server conducts preprocessing of CSI and people counting and localization. CSI preprocessing includes offset removal of amplitudes and phases for each packet, and a learning model derives a sensing result based on the training and test sample sets of preprocessed CSI.} \label{fig2}
\end{figure}

In this paper, we configure the people counting and localization process depicted in Fig. \ref{fig2}. Erroneous CSI conveyed to the server is preprocessed to remove offsets, followed by a learning model to derive a sensing result through training and test phases. The detailed CSI preprocessing procedure will be explained in Section \ref{sec4-1}. While a basic learning model is depicted in Fig. \ref{fig2}, we will adopt adaptive learning models with a convolutional neural network (CNN) in Sections \ref{sec3} and \ref{sec4-2}. Note that each task of people counting and localization is separately optimized depending on specific use of human sensing. Although we employ the same feature of input data and CNN structure, the two tasks are independently conducted by computing loss functions with different labeling. The localization task can also result in a different output form with either classification or regression, which will be specified in Section \ref{sec3}.

\section{Preliminaries}\label{sec3}

In this section, we first describe a CNN structure as a base module, which is commonly employed for adaptive learning models. As a baseline model, we introduce pre-training-based people counting and localization models to handle the situation that the characteristics of CSI samples largely depend on measurement environments. It is assumed that CSI preprocessing is applied to both the CNN module and the baseline model discussed in this section.


\begin{figure*}
	\centering
	\includegraphics[width=1.5\columnwidth]{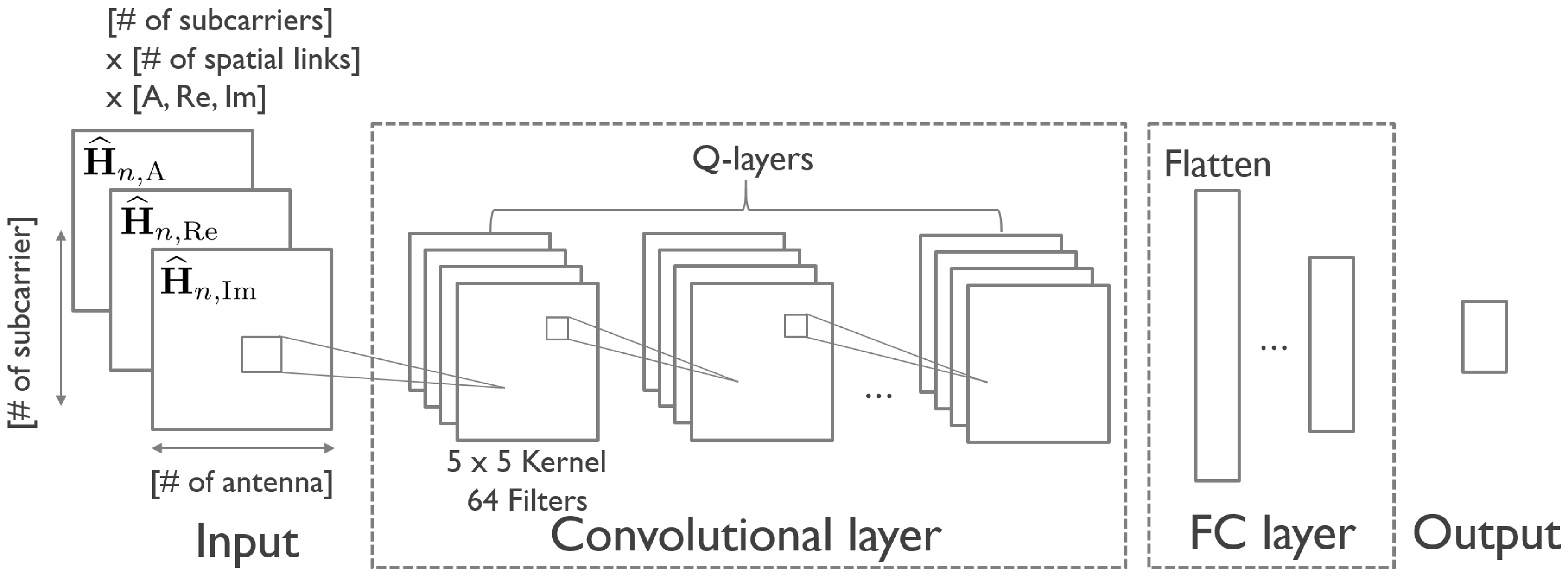}
	\caption{A basic CNN module for the proposed people counting and localization models.} \label{fig3}
\end{figure*}

\subsection{CNN Module for Classification}\label{sec3-1}

We employ a simple CNN module to operate rapidly, which is enough to achieve high classification accuracy, as shown in Section \ref{sec5-2}.  In Fig. \ref{fig3}, the preprocessed CSI matrices $ \widehat{\bH}_{n,\mathrm{A}} $, $ \widehat{\bH}_{n,\mathrm{Re}}, $ and $ \widehat{\bH}_{n,\mathrm{Im}} $ for each $ n $ are exploited as a CNN module input whose dimension is $ K\times M\times 3 $. The CNN module consists of $ Q $ convolutional layers, fully connected (FC) layers including a flatten layer, a softmax layer, and a output layer to derive a prediction result. We define the output vector of the softmax layer as
\begin{align}\label{CNN softmax output}
\hat{\bp}_n = \left[\hat{p}_{n,0},\hat{p}_{n,1},\cdots,\hat{p}_{n,C-1}\right]^\mathrm{T},
\end{align}
where $ \hat{p}_{n,c} $ is the likelihood of $ c $-th class. For people counting, each class indicates the number of people in a room. Classification can also be applied to localization by dividing a room into $ C $ sectors, and each sector is labeled by ${c \in \{0,1,\cdots, C-1\}}$. The input-output relationship of classification is expressed as
\begin{align}\label{CNN input output}
\hat{\bp}_n = f_{\text{Cla}, \boldsymbol{\Theta}}\left(\widehat{\bH}_{n,\mathrm{A}}, \widehat{\bH}_{n,\mathrm{Re}}, \widehat{\bH}_{n,\mathrm{Im}}\right),
\end{align}
where $ \boldsymbol{\Theta} $ denotes the model parameters of the CNN. Prediction is finally conducted by computing
\begin{align}\label{CNN output}
\hat{c}_n = \argmax_{c\in \{0,1,\cdots,C-1\}}\hat{p}_{n,c},
\end{align}
at the output layer.

\subsection{CNN Module for Regression}\label{sec3-2}
Classification is an effective approach for localization; however, labeling sectors is neither directly related to physical locations of people nor enough to predict precise locations. Accordingly, we also consider the CNN module for regression, and a softmax layer is not included in this module.
The input of CNN module is the same as the classification model in Section \ref{sec3-1}. The output is the location of $N_L$ people, which is defined as
\begin{align}
\hat{\bd}_n=[\hat{\bd}_{n,1}^\mathrm{T},\cdots,\hat{\bd}_{n,l}^\mathrm{T},\cdots,\hat{\bd}_{n,N_L}^\mathrm{T}]^\mathrm{T},
\end{align}
where $\hat{\bd}_{n,l}=(\hat{x}_{n,l}, \hat{y}_{n,l})^\mathrm{T}$ is the predicted location of $l$-th person. The input-output relationship of regression is given as
\begin{align}\label{CNN input output2}
\hat{\bd}_n = f_{\text{Reg}, \boldsymbol{\Theta}}\left(\widehat{\bH}_{n,\mathrm{A}}, \widehat{\bH}_{n,\mathrm{Re}}, \widehat{\bH}_{n,\mathrm{Im}}\right).
\end{align}
We will verify performance of localization using each of classification and regression models in Section \ref{sec5-2}.

\subsection{Basic Learning Process}\label{sec3-3}
We group the preprocessed CSI matrices for each phase of learning. The CSI sample set is defined as
\begin{align}\label{matrix set}
\cH_s=\left\{\left(\widehat{\bH}_{n,\mathrm{A}}, \widehat{\bH}_{n,\mathrm{Re}}, \widehat{\bH}_{n,\mathrm{Im}}\right) \big| n\in \cI_s \triangleq \cI(I_s,N_s) \right\}.
\end{align}
The received packet index set of the matrices is denoted by $ \cI(I_s,N_s) $ expressed as
\begin{align}\label{index set}
\cI(I_s,N_s)=\{n \mid I_s\leq n < I_s + N_s\},
\end{align} where $ I_s $ is the initial index of measurement, and $ N_s $ is the number of samples. If we employ the basic model in Fig. \ref{fig2}, learning process consists of simple training and test phases, and their corresponding sample sets $\cH_\text{train},\cH_\text{test}$ and index sets $\cI_\text{train},\cI_\text{test}$ can be defined. With $\cH_\text{train}$, the model parameters $ \boldsymbol{\Theta} $ is optimized to minimize a loss function, and two kinds of functions can be defined for classification and regression.

For classification, the categorical crossentropy is employed as the loss function. It measures the difference between the detected likelihood vector $ \hat{\bp}_n $ and the one-hot vector $ \bee_n=[e_{n,0},e_{n,1},\cdots,e_{n,C-1}]^\mathrm{T} $ that indicates the true value of number of people or sector index. The loss function is~computed~as
\begin{align}\label{categorical crossentropy}
\text{CE}(\cH_\text{train} ; \boldsymbol{\Theta}) = -\frac{1}{N_\text{train}}\sum_{n\in \cI_\text{train}}\sum_{c=0}^{C-1}e_{n,c}\log {\hat{p}_{n,c}}.
\end{align}
For regression, the loss function is the mean squared error (MSE) between the true location $ \bd_{n,l} $ and predicted location $ \hat{\bd}_{n,l} $, which is defined as
\begin{align}\label{mean square error}
\text{MSE}(\cH_\text{train} ; \boldsymbol{\Theta}) = \frac{1}{N_\text{train}}\sum_{n\in \cI_\text{train}}\sum_{l=1}^{N_L}\norm{\bd_{n,l}-\hat{\bd}_{n,l}}^2.
\end{align}
We define the loss function according to classification and regression as follows:
\begin{align}\label{loss function}
\text{Loss}(\cH_\text{train} ; \boldsymbol{\Theta})=\begin{cases}\text{CE}(\cH_\text{train} ; \boldsymbol{\Theta})~ &\text{for classification},\\
\text{MSE}(\cH_\text{train} ; \boldsymbol{\Theta})~&\text{for regression.}
\end{cases}
\end{align}
The optimized model is finally evaluated with the preprocessed CSI samples in the test set $\cH_\text{test}$.

\subsection{Pre-Training-Based People Counting and Localization}\label{sec3-4}
\begin{figure}
	\centering
	\includegraphics[width=0.7\columnwidth]{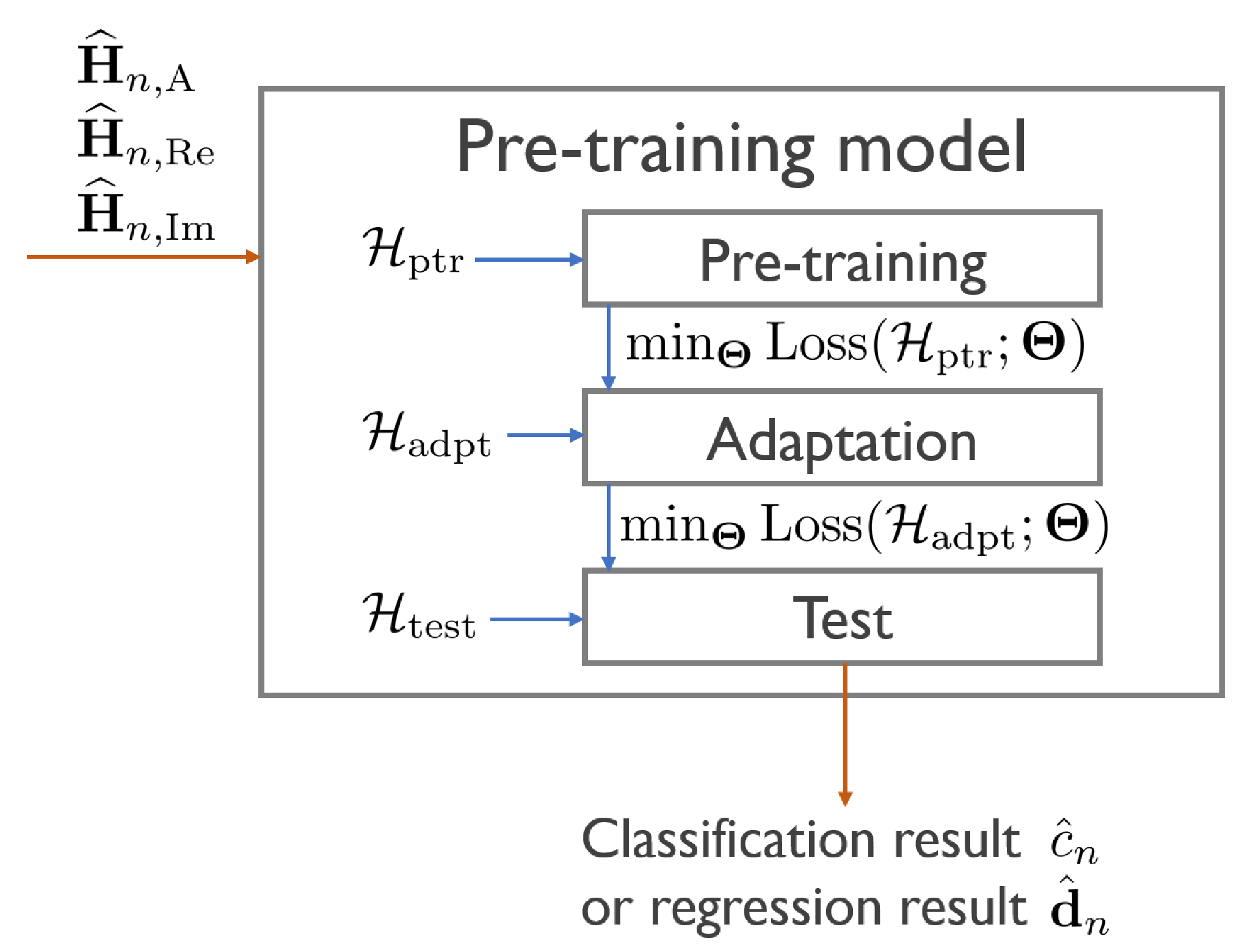}
	\caption{A pre-training-based model using preprocessed CSI as model inputs. The pre-training-based model employs a simple CNN module, whose parameters are optimized through pre-training and adaptation phases.} \label{fig4}
\end{figure}

Since the model parameters $ \boldsymbol{\Theta} $ are learned by using the training samples and employed to validate accuracy with test samples, training and test sets should be related in terms of domain and task \cite{Pan:2010}. Practically, the training set consists of outdated samples, and the test set is lately collected. Although the CSI of interest $ h(n,k,m) $ can be made consistent by maintaining the measurement environment of interest, external WiFi devices induce the interfering CSI $ h_I(n,k,m) $ \cite{Zheng:2017,Huang:2020}, which easily varies in time due to the uncontrollable change of external environments. Therefore, adaptation is required before test where adaptation and test samples are more related than~pre-training~ones.

A pre-training-based model with preprocessed CSI is depicted in Fig. \ref{fig4}.  The pre-training CSI set is denoted by $ \cH_\text{ptr} $ with the index set $ \cI_\text{ptr} $, and the adaptation set $ \cH_\text{adpt} $ consists of the samples for $ n\in\cI_\text{adpt}$. The loss function can be written~as
\begin{align}\label{pre loss function}
\text{Loss}(\cH_s ; \boldsymbol{\Theta})=\begin{cases}\text{CE}(\cH_s ; \boldsymbol{\Theta})~ &\text{for classification},\\
\text{MSE}(\cH_s ; \boldsymbol{\Theta})~&\text{for regression,}
\end{cases}
\end{align}
with equivalent definitions of \eqref{categorical crossentropy}, \eqref{mean square error}\blue{,} and \eqref{loss function} but for ${s \in \{\text{ptr},\text{adpt}\}}$.

The parameters $ \boldsymbol{\Theta} $ are first optimized with the samples in $ \cH_\text{ptr} $ during the pre-training phase with an adaptive moment estimation (ADAM) optimizer to minimize the loss function $\text{Loss}(\cH_\text{ptr} ; \boldsymbol{\Theta})$. Afterwards, regarded as an initial point of adaptation, the model parameters are adjusted by using $ \cH_\text{adpt} $. This parameter optimization during the adaptation phase is conducted with the ADAM optimizer to minimize $\text{Loss}(\cH_\text{adpt} ; \boldsymbol{\Theta})$. The pre-training-based model is evaluated with the preprocessed CSI samples in the test set $ \cH_\text{test} $. Note that the index sets do not share the same elements for each learning phase, i.e., $ {\cI_s \cap \cI_{s'} = \varnothing} $ for $ s\neq s' $ where $ s,s'\in\{\text{ptr},\text{adpt},\text{test}\} $.




\section{Proposed CSI Preprocessing and Meta-Learning-Based Scheme}\label{sec4}
In this section, we first propose the CSI preprocessing for the raw data to address the indistinguishable profiles of the input data caused by offset. Additionally, to mitigate performance degradation caused by discrepancies between training and test environments, we develop the meta-learning-based people counting and localization schemes that enable adaptive learning and generalization in diverse scenarios.

\subsection{CSI Preprocessing}\label{sec4-1}

\begin{figure}
	\centering
	\includegraphics[width=0.95\columnwidth]{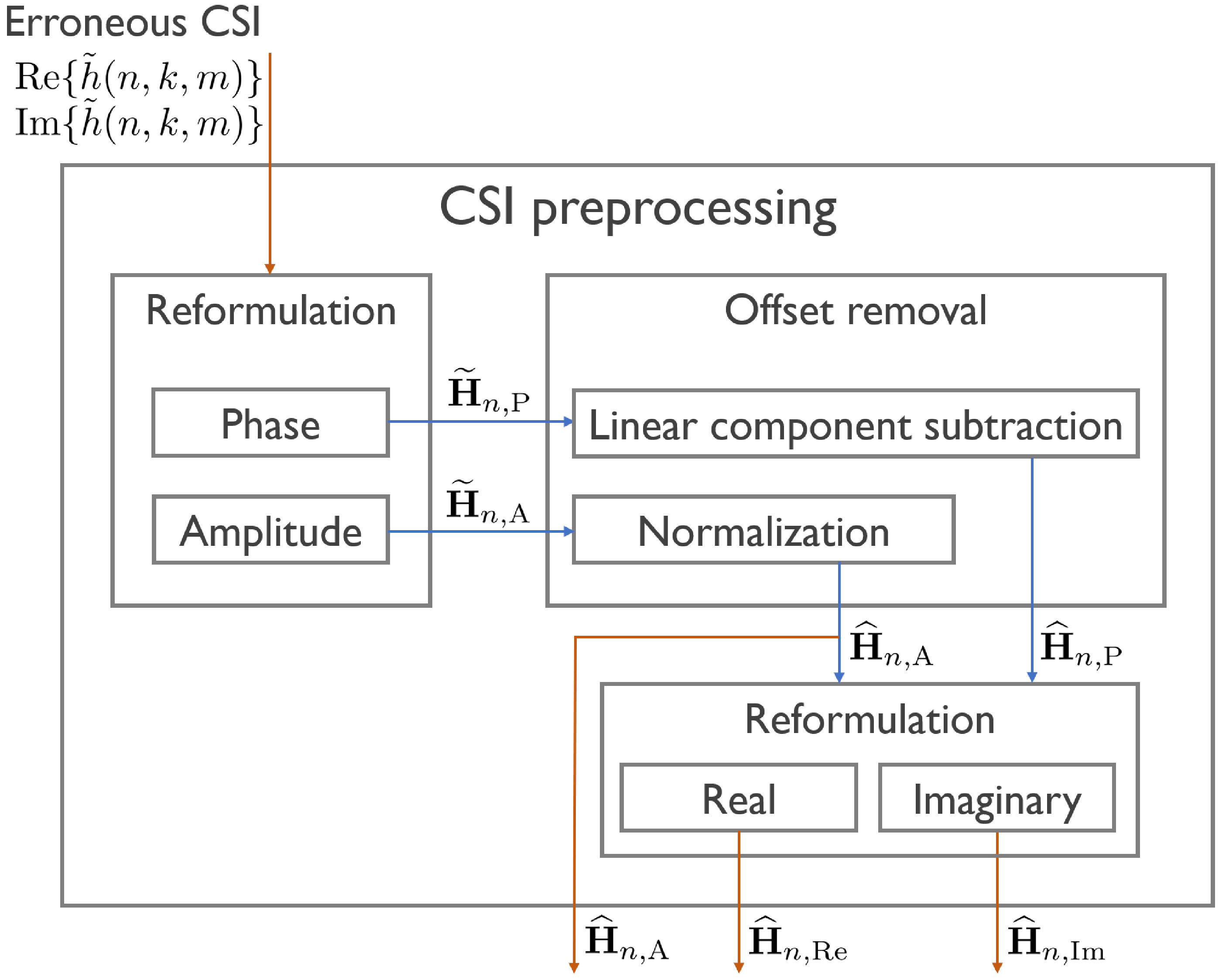}
	\caption{Procedure of CSI preprocessing. From erroneous CSI, a distinct feature of a given environment is extracted by reducing randomly generated offsets.} \label{fig5}
\end{figure}

We propose preprocessing of CSI for each packet to improve learning efficiency. Raw data of CSI including offsets leads to indistinguishable profiles of real and imaginary values of CSI over different classes, which degrades learning performance. To derive distinct features that are consistent for the same class but different from other classes, we propose effective preprocessing for learning models, as depicted in Fig. \ref{fig5}.

When the $ n $-th packet is captured at the receiver, the measured CSI conveyed to the server is expressed as $ \Re\left\{\tilde{h}(n,k,m)\right\} $ and $ \Im\left\{\tilde{h}(n,k,m)\right\} $ for $ k\in \{1,2,\cdots,K\} $ and $ m\in \{1,2,\cdots,M\} $. For preprocessing, we construct CSI as a $ K\times M $ complex-valued matrix, whose $ (k , m) $-th element is written as
\begin{align}\label{CSI matrix}
\left[\widetilde{\bH}_{n}\right]_{(k,m)} =\Re\left\{\tilde{h}(n,k,m)\right\} + j\Im\left\{\tilde{h}(n,k,m)\right\}.
\end{align}
The CSI matrix $ \widetilde{\bH}_{n} $ can be separated as its amplitudes and phases for all elements. We define these separated matrices as $ \widetilde{\bH}_{n,\mathrm{A}} $ and $ \widetilde{\bH}_{n,\mathrm{P}} $ where each element is expressed as
\begin{align}\label{CSI amplitude phase}
\left[\widetilde{\bH}_{n,\mathrm{A}}\right]_{(k,m)} &= \left\lvert\left[\widetilde{\bH}_n\right]_{(k,m)}\right\rvert,\\
\left[\widetilde{\bH}_{n,\mathrm{P}}\right]_{(k,m)} &= \angle{\left(\left[\widetilde{\bH}_n\right]_{(k,m)}\right)}.
\end{align}

Phase offset removal is a general procedure for human sensing using CSI. As seen in \eqref{erroneous CSI}, there are several factors causing the phase offsets, which vary with time and frequency. Most of previous studies handled the issue by using only the amplitudes of CSI \cite{Wang:2017a,Jiang:2020,Wang:2017b}. However, to keep the information of signal propagation in \eqref{propagation delay}, we employ both the phases and amplitudes of CSI. Since a general DNN benefits from a large number of training samples, we are also motivated to conserve the number of CSI samples without stacking over packets for filtering.

We conduct phase offset removal over subcarriers for each packet. Recalling the exponent of phase offsets $\left(\epsilon_{b}+\epsilon_{s}(n)\right)k+\epsilon_c(n)$ in \eqref{erroneous CSI}, it is observed that the phase rotation occurs by the slope $ {\epsilon_{b}+\epsilon_{s}(n)} $ proportional to the subcarrier index $ k $ and the constant $ \epsilon_c(n) $ for each packet [27], [28]. Because this random phase rotation varies with packets, for each of $ M $ spatial links, we first conduct linear regression for unwrapped phases of CSI $ [\widetilde{\bH}_{n,\mathrm{P}}]_{(:,m)} $ over subcarriers and subtract the linear component from the raw phases.
Specifically, we first define a $ K\times 2 $ auxiliary matrix for linear regression as
\begin{align}\label{Auxiliary matrix}
\bX = \begin{bmatrix}
1&1&\cdots&1\\1&2&\cdots&K
\end{bmatrix}^\mathrm{T}.
\end{align}
We then compute the parameter vector including the regression component as
\begin{align}\label{parameter vector}
\boldsymbol{\beta}_{n,m}=\bX^\dagger \left[\widetilde{\bH}_{n,\mathrm{P}}\right]_{(:,m)}.
\end{align}
Removing the linear component with $ \boldsymbol{\beta}_{n,m} $ for each spatial link $ m $, the column vector of phase matrix is derived as
\begin{align}
\left[\overline{\bH}_{n,\mathrm{P}}\right]_{(:,m)}= \left[\widetilde{\bH}_{n,\mathrm{P}}\right]_{(:,m)}-\bX\boldsymbol{\beta}_{n,m}.
\end{align}
There still exists the constant offset for all subcarriers, which can be reduced by subtracting the phase at the reference subcarrier $ k_0 $ to all phases of the $ K $ subcarriers. The phase offset removal is finally conducted as
\begin{align}\label{CSI phase offset}
\widehat{\bH}_{n,\mathrm{P}} = \overline{\bH}_{n,\mathrm{P}} - \mathbf{1}_K\left[\overline{\bH}_{n,\mathrm{P}}\right]_{(k_0,:)}.
\end{align}

In addition to phase offsets, there also exist amplitude offsets by uncertainty in power control \cite{Xie:2019}. While previous works, e.g., \cite{Zhao:2019,Choi:2021}, conducted filtering process for amplitudes of CSI of consecutive packets, we normalize the amplitudes over subcarriers for each packet and spatial link, written as
\begin{align}\label{CSI amplitude offset}
\left[\widehat{\bH}_{n,\mathrm{A}}\right]_{(:,m)} = 
\frac{\left[\widetilde{\bH}_{n,\mathrm{A}}\right]_{(:,m)}}{\frac{1}{K}\sum_{k=1}^K\left[\widetilde{\bH}_{n,\mathrm{A}}\right]_{(k,m)}}.
\end{align}
Based on $ \widehat{\bH}_{n,\mathrm{A}} $ and $ \widehat{\bH}_{n,\mathrm{P}} $, we construct a complex-valued preprocessed CSI matrix and separate it into real and imaginary parts to derive two real-valued matrices, whose elements are~expressed~as
\begin{align}\label{CSI processed}
\left[\widehat{\bH}_{n,\mathrm{Re}}\right]_{(k,m)} = \Re\left\{{\left[\widehat{\bH}_{n,\mathrm{A}}\right]_{(k,m)} e^{j \left[\widehat{\bH}_{n,\mathrm{P}}\right]_{(k,m)}}}\right\},\\
\left[\widehat{\bH}_{n,\mathrm{Im}}\right]_{(k,m)} = \Im\left\{{\left[\widehat{\bH}_{n,\mathrm{A}}\right]_{(k,m)} e^{j \left[\widehat{\bH}_{n,\mathrm{P}}\right]_{(k,m)}}}\right\}.
\end{align}
For each packet, we employ these preprocessed matrices $ \widehat{\bH}_{n,\mathrm{A}} $, $ \widehat{\bH}_{n,\mathrm{Re}}, $ and $ \widehat{\bH}_{n,\mathrm{Im}} $ as CNN model inputs.\footnote{The real, imaginary, and amplitude of channels are used to the CNN inputs for the performance improvements in \cite{Ahmet:2020}.} Different from existing methods, there are no additional denoising and filtering processes that require CSI stacking over packets, which enables to exploit a large number of CSI samples separately by packet. To verify the impact of CSI preprocessing, the detailed performance comparison with the raw CSI data is presented in Section \ref{sec5-2}.

\subsection{Meta-Learning-Based People Counting and Localization Models}\label{sec4-2}

\begin{figure}
	\centering
	\includegraphics[width=0.95\columnwidth]{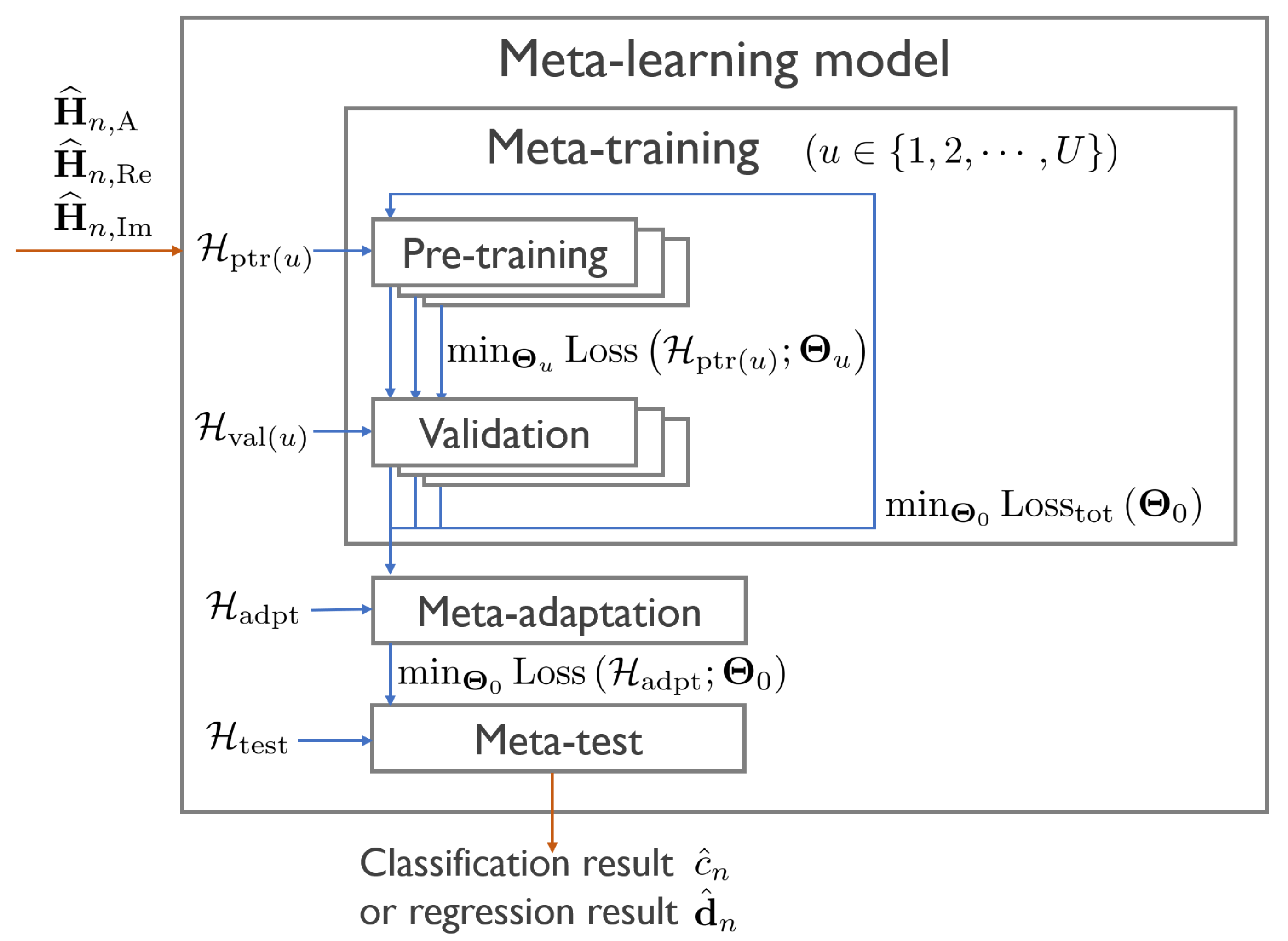}
	\caption{The proposed meta-learning-based people counting and localization models. The meta-training phase is to learn the way how well the model is adaptive to each of various environments. The meta-adaptation phase exploits only a few adaptation samples from a new environment.} \label{fig6}
\end{figure}

Pre-trained models explained in Section \ref{sec3-4} are used as a straightforward approach to people counting and localization. However, when the test environment differs from the training environment, these pre-trained models may suffer from considerable performance loss. This is because the model parameters can still be dependent on the training environment even after adaptation. To mitigate this issue, we propose the meta-learning-based people counting and localization models employing MAML for better initialization and adjustment of model parameters in meta-adaptation and meta-test phases. As depicted in Fig. \ref{fig6}, the proposed learning model consists of meta-training phase that includes multiple pre-training and validation processes, meta-adaptation phase, and meta-test phase. The proposed model equivalently adopts the CNN structure and the CSI preprocessing explained in Sections \ref{sec3-1}, \ref{sec3-2}, and \ref{sec4-1}. However, during the meta-training phase, the model parameters learn a general way the model can be easily adapted to any measurement environments. The parameters are consequently optimized into more generalized ones and can become adequately adjusted for the meta-adaptation and meta-test environment.

For the CSI grouping in the meta-learning-based model with the definition of CSI sample set in \eqref{matrix set}, several pre-training sets $ \cH_{\text{ptr}(u)} $ for $ {u \in \{1,2,\cdots,U\}} $ are chosen as $ U $ specific tasks. Each of the sets includes the CSI samples corresponding to the index set $ \cI_{\text{ptr}(u)} $. Similarly, we define the validation sets $ \cH_{\text{val}(u)} $ and the corresponding index sets $ \cI_{\text{val}(u)} $ for $ u \in \{1,2,\cdots,U\} $. The CSI sample sets for meta-adaptation and meta-test phases are denoted by $ \cH_\text{adpt} $ and $ \cH_\text{test} $ same as in Section~\ref{sec3-4}. Meanwhile, similar to \eqref{pre loss function}, the loss function in the meta-learning-based model can be expressed as
\begin{align}\label{inner loss function}
\mathrm{Loss}\left(\cH_s;\boldsymbol{\Theta}_t\right)=\begin{cases}\text{CE}(\cH_s ; \boldsymbol{\Theta}_t)~ &\text{for classification},\\
\text{MSE}(\cH_s ; \boldsymbol{\Theta}_t)~&\text{for regression,}
\end{cases}
\end{align}
for $ {s\in \bigcup_{u=1}^U \{\text{ptr}(u),\text{val}(u)\}\cup\{\text{adpt}\}, t\in \{0,1,\cdots,U\}} $.

During the meta-training phase, the proposed model optimizes the meta-learner parameters $ \boldsymbol{\Theta}_\mathrm{0} $ such that the task-specific parameters $ \boldsymbol{\Theta}_u $ can become appropriately adaptive for each $u \in \{1,2,\cdots,U\}$. As in Fig. \ref{fig6}, the meta-training phase has two updates: inner-loop update and outer-loop update. For the first iteration, the model parameters $ \boldsymbol{\Theta}_u $ are set to the randomly initialized parameters $ \boldsymbol{\Theta}_\mathrm{0} $. In the inner-loop update, the pre-training is conducted with  $ \cH_{\text{ptr}(u)} $ to minimize the loss function $ \mathrm{Loss}\left(\cH_{\text{ptr}(u)};\boldsymbol{\Theta}_u\right) $ in \eqref{inner loss function} where a stochastic gradient descent (SGD) optimizer is utilized
\begin{align}
\boldsymbol{\Theta}_u\leftarrow \boldsymbol{\Theta}_u-\alpha \nabla_{\boldsymbol{\Theta}_u} \mathrm{Loss}\left(\cH_{\text{ptr}(u)};\boldsymbol{\Theta}_u\right),
\end{align}
with the inner-loop rate $\alpha$.

In the outer-loop update, based on the task-specific parameters $ \boldsymbol{\Theta}_u$ optimized for each of $ U $ tasks, $ \boldsymbol{\Theta}_\mathrm{0} $ is updated with the ADAM optimizer to minimize the total loss function. It is defined as
\begin{align}\label{meta learning loss function}
\text{Loss}_\text{tot}(\boldsymbol{\Theta}_0) =\sum_{u=1}^{U} \text{Loss}\left(\cH_{\text{val}(u)};\boldsymbol{\Theta}_u\right),
\end{align}
with the outer-loop rate $\beta$ by evaluating the model with the validation samples in $ \cH_{\text{val}(u)} $.
The updated parameters $ \boldsymbol{\Theta}_\mathrm{0} $ are employed as new initial values of $ \boldsymbol{\Theta}_u $ for the next iteration, and the optimization of $ \boldsymbol{\Theta}_\mathrm{0} $ is progressed through the following pre-training and validation phases. By conducting the process with a few iterations, $ \boldsymbol{\Theta}_\mathrm{0} $ is gradually optimized.

After the meta-training phase, the initial model parameters are adjusted to minimize $ \mathrm{Loss}\left(\cH_\text{adpt};\boldsymbol{\Theta}_\mathrm{0}\right) $ during the meta-adaptation phase with the SGD optimizer as
\begin{align}
\boldsymbol{\Theta}_0\leftarrow \boldsymbol{\Theta}_0-\alpha \nabla_{\boldsymbol{\Theta}_0} \mathrm{Loss}\left(\cH_{\text{adpt}};\boldsymbol{\Theta}_0\right).
\end{align}
This adaptive model is finally evaluated with the CSI samples in $ \cH_{\text{test}} $ during the meta-test phase. As in Section \ref{sec3-4}, the defined index sets among different learning phases and tasks are disjoint each other, i.e., $ {\cI_s \cap \cI_{s'} = \varnothing} $ for $ s\neq s' $ where ${ s,s'\in\bigcup_{u=1}^U \{\text{ptr}(u),\text{val}(u)\}\cup\{\text{adpt},\text{test}\}} $.

\section{Complexity Analysis and Experimental Results}\label{sec5}
 In this section, we first analyze the computational complexity of the proposed scheme and the CSI preprocessing. Then, we evaluate the performance of the people counting and localization models in the experimental results.

\subsection{Complexity Analysis}\label{sec5-1}
In this subsection, we analyze the computational complexity of the meta-learning-based model, the pre-trained model, the transfer-learning-based model, which are denoted as Meta, Pre, and TL, respectively, and the CSI preprocessing with the Big-O notation \cite{Hunger:2005}. The computational complexity of Meta is divided into three phases, which are training, adaptation, and testing phases. In the training phase, the computational complexity is given by \cite{Mizutani:2001, Taghavi:2019, Kim:2021, Kim:2023}
\begin{align}
&C_\text{Meta-train}\notag \\&=\mathcal{O}\Bigg(N_\text{epoch}N_\text{meta-train}\Bigg(\sum_{q=1}^{Q}N_{\text{ker}}C_{\text{in},q}C_{\text{out},q}+\sum_{i=1}^LD_iD_{i+1}\notag\\ &+D_{L+1}C\Bigg)\Bigg)\notag\\
&\stackrel{(a)}{=}\mathcal{O}\big(N_\text{epoch}N_\text{meta-train}\big(N_\text{ker}\big(C_0N_f+(Q-1)N_f^2\big)+C_QN_d\notag\\
&+(L-1)N_d^2+N_dC\big)\big)\notag \\
&\stackrel{(b)}{=}\mathcal{O}\left(N_\text{epoch}N_\text{meta-train}\left(N_\text{ker}QN_f^2+LN_d^2+N_dC\right)\right)\notag\\
&\stackrel{(c)}{=}\mathcal{O}\left(N_\text{epoch}N_\text{meta-train}\left(N_\text{ker}QN_f^2+LN_d^2\right)\right),
\end{align} where $N_\text{epoch}$ is the number of epoch, $N_\text{meta-train}$ is the number of meta-training samples, $Q$ is the number of the convolutional layers, $N_\text{ker}$ is the size of kernel, $C_{\text{in},q}$ is the size of the $q$-th kernel input, $C_{\text{out},q}$ is the size of the $q$-th kernel output, $L$ is the number of FC layers, $D_i$ is the number of the $i$-th nodes, $N_f$ is the number of filters, $C_Q$ is the number of nodes in the flatten layer, and $N_d$ is the number of nodes in FC layers. Note that (a) is from $C_{\text{in},1}=C_0$ where $C_0=3$ since we use amplitude, real, and imaginary matrices as the inputs, $C_{\text{in},q}=N_f$ for $q>1$, $C_{\text{out},q}=N_f$, $D_1=C_Q$, and $D_i=N_d$, (b) comes from $N_f\gg C_0$ and $C_Q\simeq N_d$, and (c) is derived by $N_d\gg C$. Similar to the above derivation, the complexity of adaptation phase with the number of gradient steps $N_\text{gr}$ and the number of adaptation samples $N_\text{adpt}$ is given as
\begin{align}
C_\text{Meta-adaptaion}{=}\mathcal{O}\left(N_\text{gr}N_\text{adpt}\left(N_\text{ker}QN_f^2+LN_d^2\right)\right),
\end{align} and the complexity of testing phase is given by 
\begin{align}
C_\text{Meta-test}{=}\mathcal{O}(N_\text{test}(N_\text{ker}QN_f^2+LN_d^2)).    
\end{align} The total complexity of Meta then becomes
\begin{align}
C_\text{Meta}{=}&\mathcal{O}\big((N_\text{epoch}N_\text{meta-train}+N_\text{gr}N_\text{adpt}+N_\text{test})\notag \\
&\cdot\big(N_\text{ker}QN_f^2+LN_d^2\big)\big).
\end{align}

We also analyze the computational complexity of Pre and TL, which are our benchmarks in Section \ref{sec5-2}. In the training phase, the complexity of Pre can be obtained by
\begin{align}
C_\text{Pre-train}{=}\mathcal{O}\left(N_\text{epoch}N_\text{pre-train}\left(N_\text{ker}QN_f^2+LN_d^2\right)\right),
\end{align} where $N_\text{pre-train}$ is the number of training samples of Pre. Then, the total complexity of Pre is given by
\begin{align}
C_\text{Pre}{=}&\mathcal{O}\big((N_\text{epoch}N_\text{pre-train}+N_\text{gr}N_\text{adpt}+N_\text{test})\notag \\
&\cdot\big(N_\text{ker}QN_f^2+LN_d^2\big)\big).
\end{align}

In the TL method, it consists of three phases, which are training, fine-tuning, and testing phases. The complexity in training phase of TL can be expressed as
\begin{align}
C_\text{TL-train}{=}\mathcal{O}\left(N_\text{epoch}N_\text{tl-train}\left(N_\text{ker}QN_f^2+LN_d^2\right)\right),
\end{align} where $N_\text{tl-train}$ is the number of training samples of TL. In the fine-tuning phase, we assume that only the FC layers in the CNN module are fine-tuned in the TL framework. Therefore, the complexity in fine-tuning phase is given by
\begin{align}
C_\text{TL-fine}{=}\mathcal{O}\left(N_\text{gr}N_\text{adpt}LN_d^2\right).
\end{align}
Considering the complexity in testing phase, the total complexity of TL is finally given as
\begin{align}
C_\text{TL}{=}&\mathcal{O}\big((N_\text{epoch}N_\text{tl-train}+N_\text{test})\left(N_\text{ker}QN_f^2+LN_d^2\right)\notag \\&+N_\text{gr}N_\text{adpt}LN_d^2\big).
\end{align}

It is clearly observed that the total complexities of Meta, Pre, and TL are almost the same since $N_\text{epoch}N_\text{train} \gg N_\text{gr}N_\text{adpt}>N_\text{test}$. If we assume the number of training samples for each scheme is the same, the total complexities of all schemes can be approximated as $\mathcal{O}\left(N_\text{epoch}N_\text{train}\left(N_\text{ker}QN_f^2+LN_d^2\right)\right)$.

Moreover, we analyze the complexity of the CSI preprocessing. Since the operations in (\ref{Auxiliary matrix})-(\ref{CSI amplitude offset}) scale linearly with $K$, the total complexity across $M$ antennas and $N_\text{pck}$ packets becomes $\mathcal{O}(N_\text{pck}MK)$. The complexity of CSI preprocessing is much smaller than that of NN frameworks since $N_\text{epoch}N_\text{train}N_f^2\gg N_\text{pck}MK$. The computational complexities of Meta, Pre, TL, and the CSI preprocessing are summarized in Table \ref{table1}.

\begin{table*}[t!]
	\centering
	\renewcommand{\arraystretch}{1.25}
	\captionsetup{justification=centering}
	\captionsetup{labelsep=newline}
	\caption{Computational complexities of Meta, Pre, TL, and CSI preprocessing}
	\resizebox{\linewidth}{!}{%
		\begin{tabular}{l l l l}
			\toprule
			\text{Scheme} & Phase & \text{Complexity} & \text{Total complexity} \\
			\hline
			\multirow{3}{*}{Meta}& \multicolumn{1}{l}{Train} & \multicolumn{1}{l}{$\mathcal{O}\left(N_\text{epoch}N_\text{meta-train}\left(N_\text{ker}QN_f^2+LN_d^2\right)\right)$} & \multirow{3}{*}{$\mathcal{O}\left((N_\text{epoch}N_\text{meta-train}+N_\text{gr}N_\text{adpt}+N_\text{test})\left(N_\text{ker}QN_f^2+LN_d^2\right)\right)$} \\ 
			& \multicolumn{1}{l}{Adaptation} & \multicolumn{1}{l}{$\mathcal{O}\left(N_\text{gr}N_\text{adpt}\left(N_\text{ker}QN_f^2+LN_d^2\right)\right)$} & \\  
			& \multicolumn{1}{l}{Test} & \multicolumn{1}{l}{$\mathcal{O}\left(N_\text{test}\left(N_\text{ker}QN_f^2+LN_d^2\right)\right)$} & \\ \hline			
			\multirow{3}{*}{Pre}& \multicolumn{1}{l}{Train} & \multicolumn{1}{l}{$\mathcal{O}\left(N_\text{epoch}N_\text{pre-train}\left(N_\text{ker}QN_f^2+LN_d^2\right)\right)$} & \multirow{3}{*}{$\mathcal{O}\left((N_\text{epoch}N_\text{pre-train}+N_\text{gr}N_\text{adpt}+N_\text{test})\left(N_\text{ker}QN_f^2+LN_d^2\right)\right)$} \\ 
			& \multicolumn{1}{l}{Adaptation} & \multicolumn{1}{l}{$\mathcal{O}\left(N_\text{gr}N_\text{adpt}\left(N_\text{ker}QN_f^2+LN_d^2\right)\right)$} & \\  
			& \multicolumn{1}{l}{Test} & \multicolumn{1}{l}{$\mathcal{O}\left(N_\text{test}\left(N_\text{ker}QN_f^2+LN_d^2\right)\right)$} & \\ \hline
			\multirow{3}{*}{TL}& \multicolumn{1}{l}{Train} & \multicolumn{1}{l}{$\mathcal{O}\left(N_\text{epoch}N_\text{tl-train}\left(N_\text{ker}QN_f^2+LN_d^2\right)\right)$} & \multirow{3}{*}{$\mathcal{O}\left((N_\text{epoch}N_\text{tl-train}+N_\text{test})\left(N_\text{ker}QN_f^2+LN_d^2\right)+N_\text{gr}N_\text{adpt}LN_d^2\right)$} \\ 
			& \multicolumn{1}{l}{Fine-tune} & \multicolumn{1}{l}{$\mathcal{O}\left(N_\text{gr}N_\text{adpt}LN_d^2\right)$} & \\  
			& \multicolumn{1}{l}{Test} & \multicolumn{1}{l}{$\mathcal{O}\left(N_\text{test}\left(N_\text{ker}QN_f^2+LN_d^2\right)\right)$} & \\ \hline		
			\multirow{1}{*}{CSI preprocessing} & \multicolumn{1}{l}{-} &\multicolumn{1}{l}{$\mathcal{O}\left(N_\text{pck}MK\right)$} & \multirow{1}{*}{-} \\ 		
			\bottomrule
	\end{tabular}}
	{\label{table1}}
\end{table*}

\subsection{Experimental Results}\label{sec5-2}
\begin{figure}[t]
	\centering
	\begin{subfigure}[c]{0.5\textwidth}
		\centering
	  $\vcenter{\hbox{\includegraphics[width=0.4\textwidth]{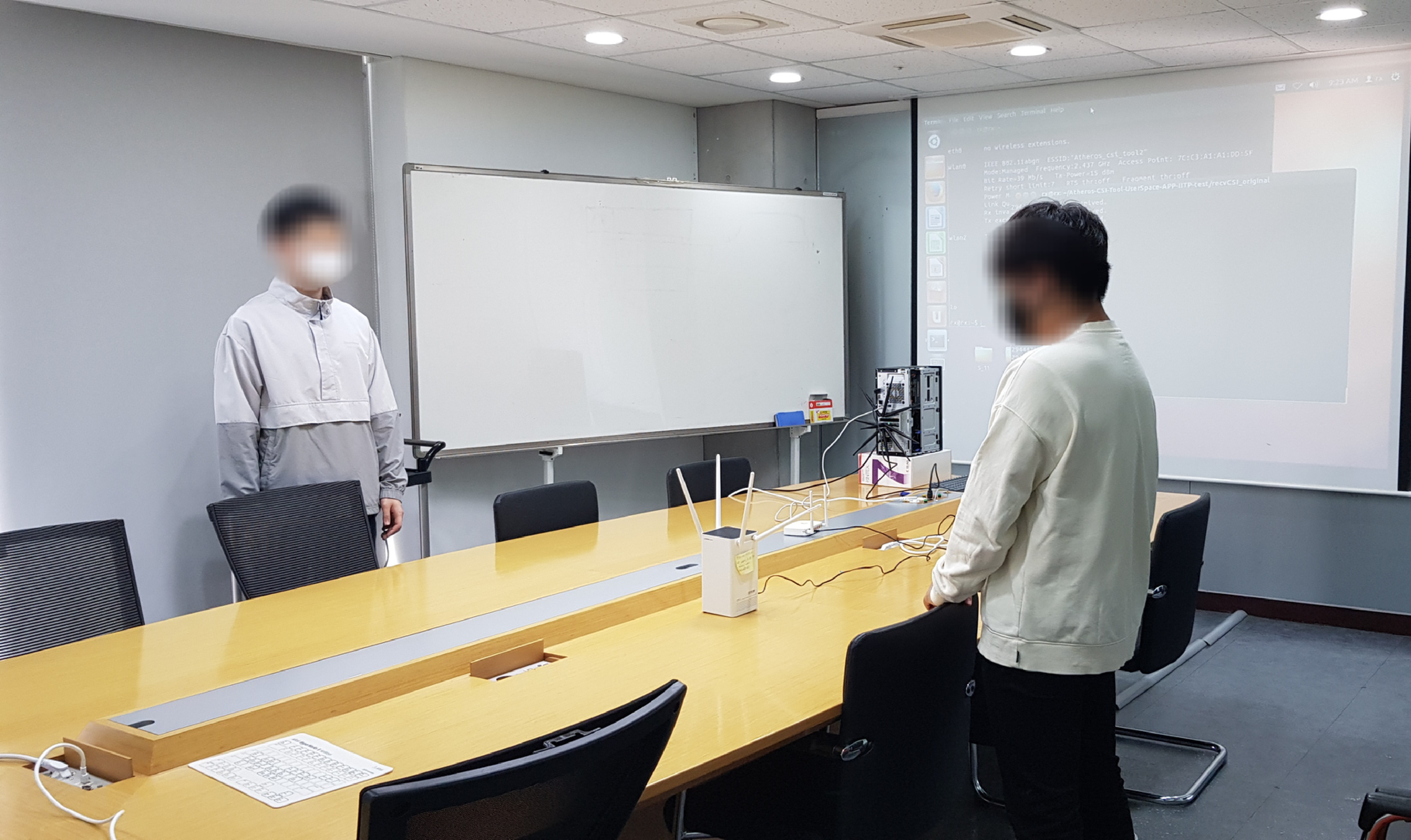}}}$%
		\hspace{0.5em}
		$\vcenter{\hbox{\includegraphics[width=0.5\textwidth]{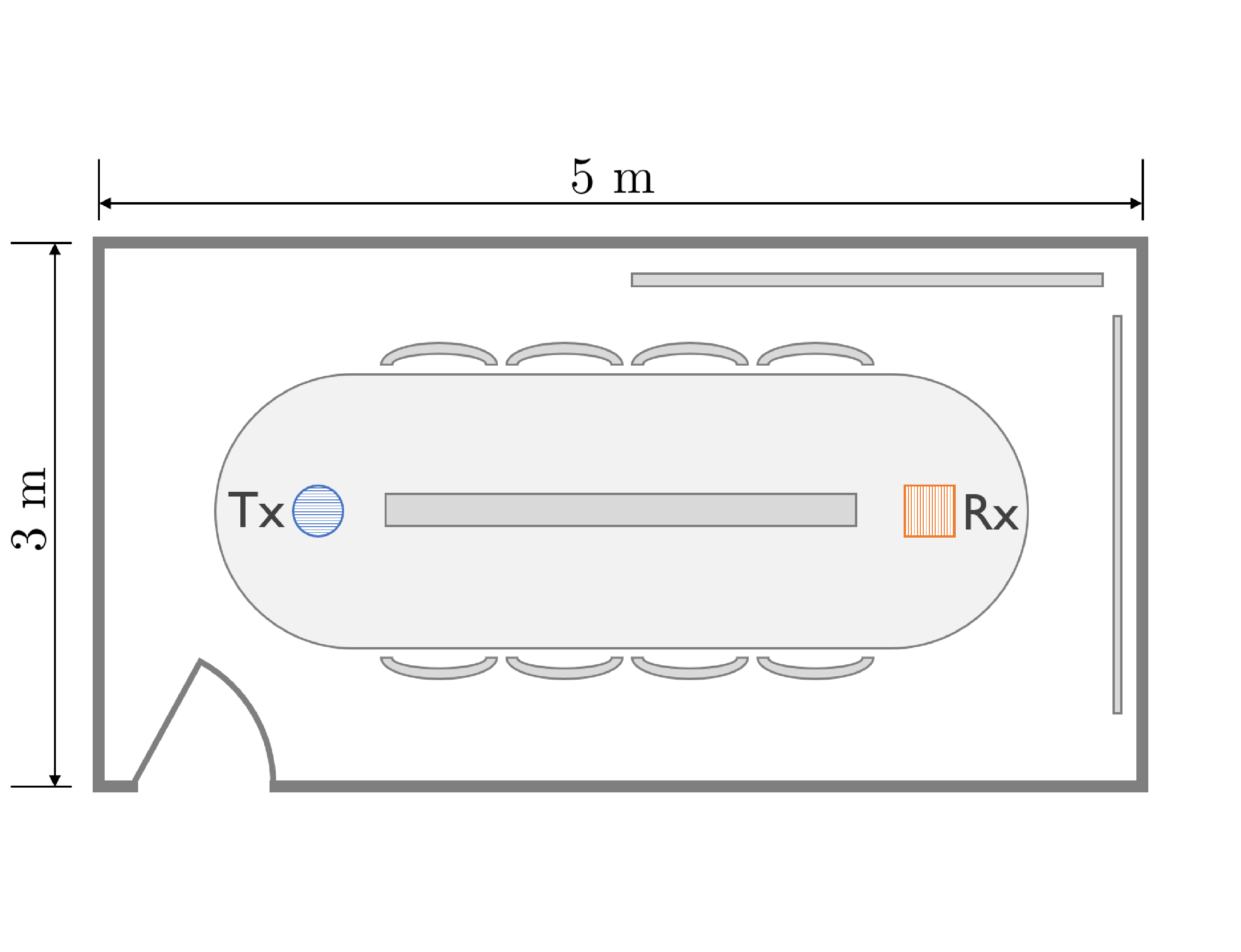}}}$
		\caption{Small room.}
	\vspace{0.5em}
		\label{fig7:first}
	\end{subfigure}
	\begin{subfigure}[c]{0.5\textwidth}
		\centering
		$\vcenter{\hbox{\includegraphics[width=0.4\textwidth]{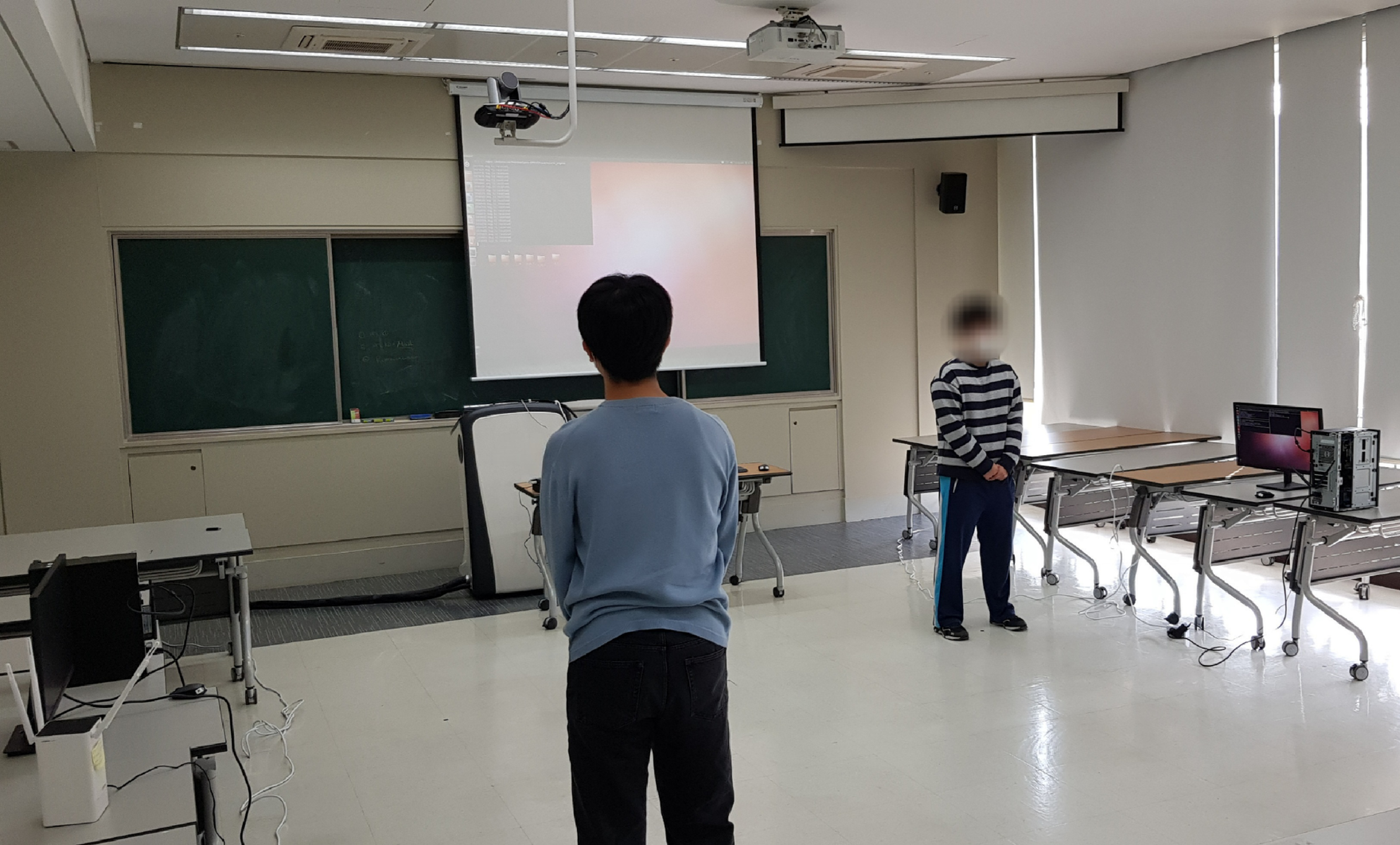}}}$%
		\hspace{0.5em}
		$\vcenter{\hbox{\includegraphics[width=0.5\textwidth]{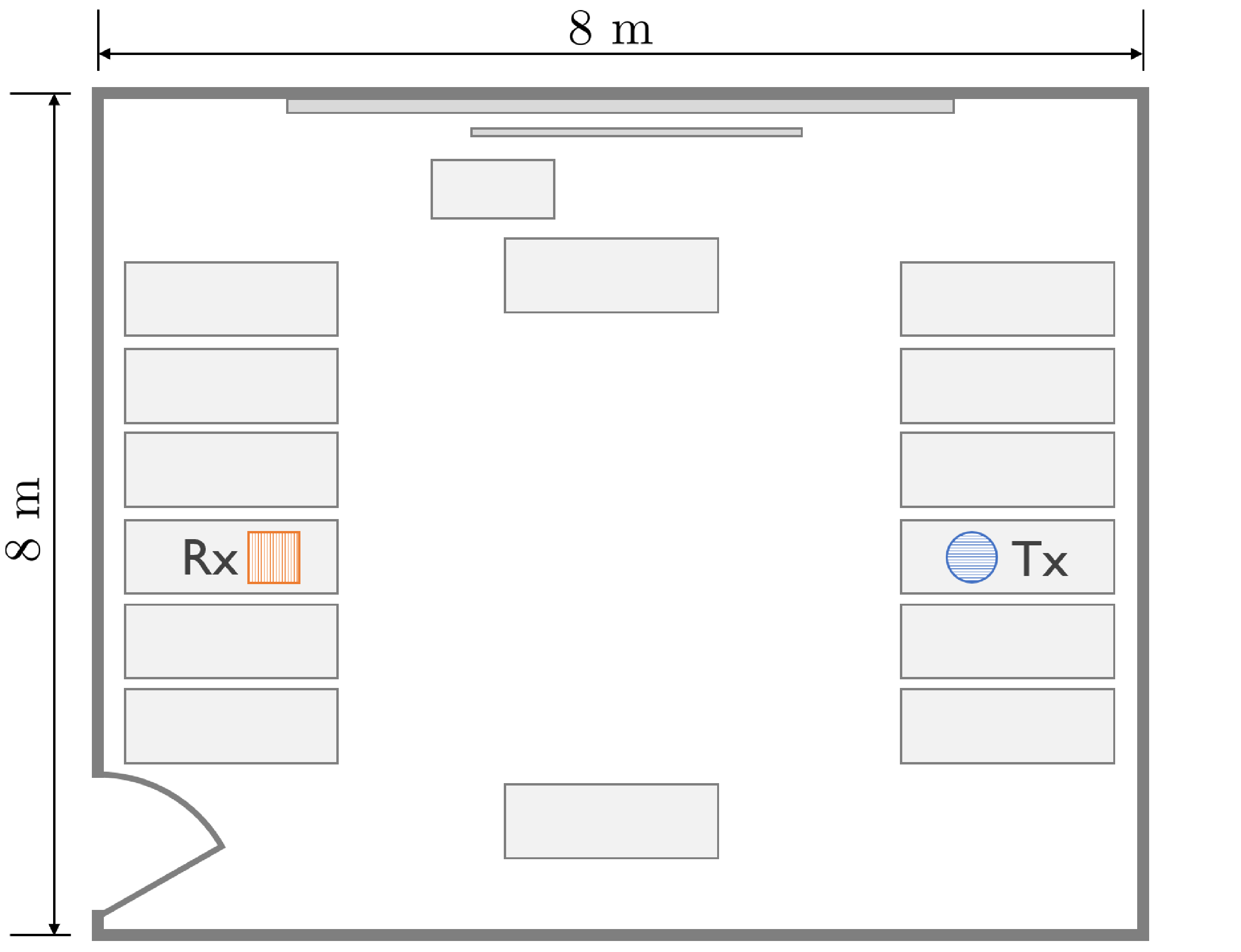}}}$
		\caption{Large room.}
	\vspace{0.5em}
		\label{fig7:second}
	\end{subfigure}
	\begin{subfigure}[c]{0.5\textwidth}
		\centering
		$\vcenter{\hbox{\includegraphics[width=0.4\textwidth]{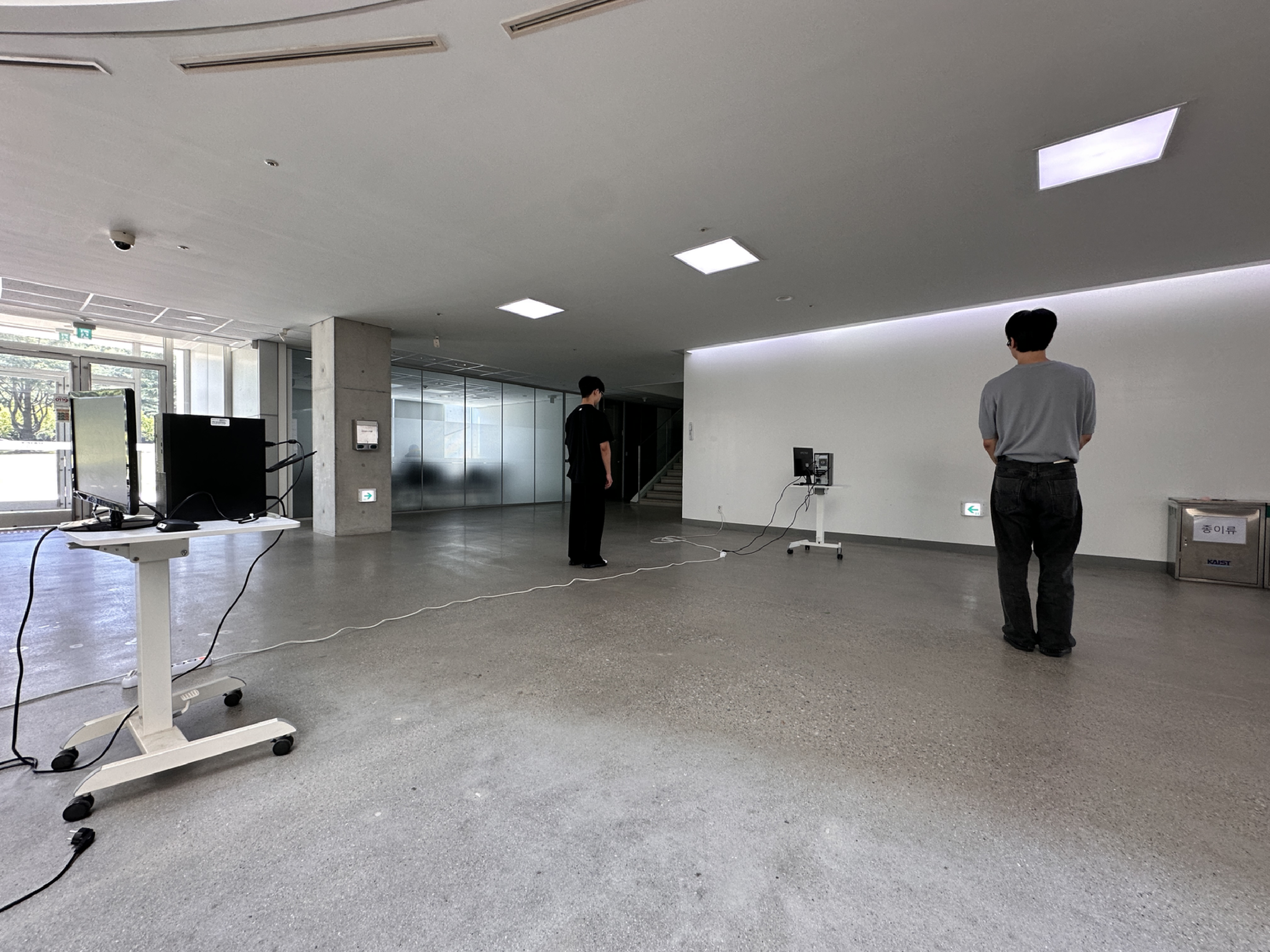}}}$%
		\hspace{0.5em}
		$\vcenter{\hbox{\includegraphics[width=0.5\textwidth]{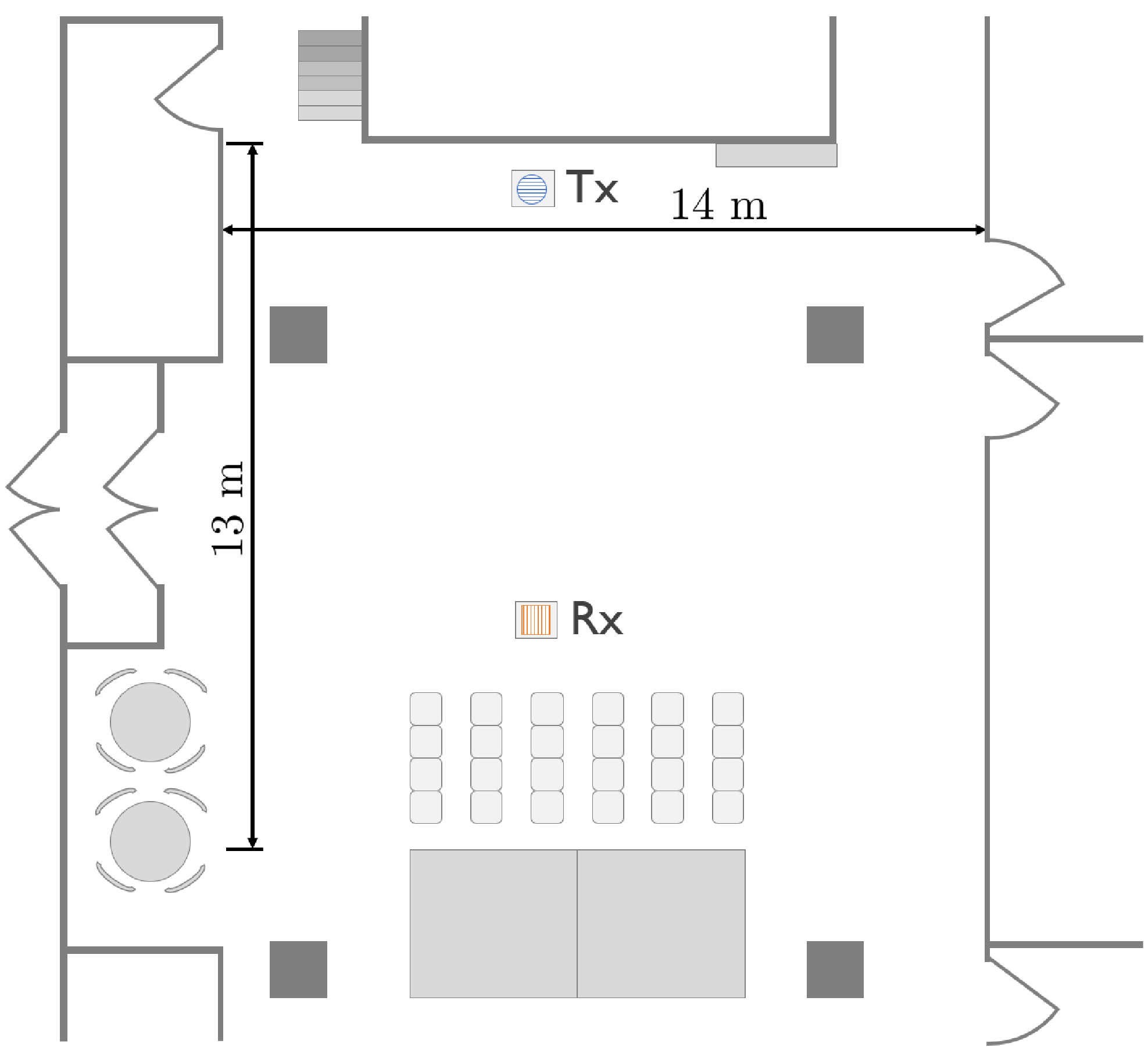}}}$
		\caption{Open space.}
		\label{fig7:third}
	\end{subfigure}
	\caption{Experimental environments for CSI measurement.}
	\label{fig7:figures}
\end{figure}

CSI measurement is conducted in two rooms, i.e., a small room ($3$ m $\times$ $5$ m) and a large room ($8$ m $\times$ $8$ m), and an open space (larger than $14$ m $\times$ $13$ m) depicted in Fig. \ref{fig7:figures}. We use the WiFi NIC (Atheros AR$9380$) with the carrier frequency $f_c=2.437$ GHz, sampling frequency $f_s=1$ kHz, number of subcarriers $K=52$, and number of spatial links $M=9$ with transceivers with three antennas $ (M_\mathrm{t}=M_\mathrm{r}=3) $. With the measured CSI, the people counting and localization models use the NVIDIA Quadro RTX $8000$ GPU and TensorFlow $2.1$ in the experimental results. For the CNN model, we use $Q=5$ convolutional layers consisting of $N_f=64$ filters with a kernel size of $5\times5$. The input dimension for CNN is $K\times M\times 3=52 \times 9 \times 3$. For the FC layers in CNN module, we use $L_\text{peo}=2$ layers with $N_d=256$ nodes for the people counting model and $L_\text{loc}=4$ layers with $N_d=256$ nodes for the localization model. In the MAML algorithm, we set the number of epochs $N_\text{epoch}=20$, the batch size $V=64$, the inner-loop rate $\alpha=10^{-1}$, and the outer-loop rate $\beta=10^{-4}$. All system parameters are summarized in Table \ref{table2}.

 In this paper, we use accuracy as a performance metric for classification, defined as
\begin{align}
\text{Accuracy} = \frac{1}{N_\text{test} }\sum_{n\in \cI_\text{test}}\delta_{\hat{c}_n,c_n},
\end{align}
where $ \delta_{\hat{c}_n,c_n} $ is the Kronecker delta function to count the number of correct predictions as $ \hat{c}_n=c_n $ for the test samples in $ \cH_\text{test} $.
Also, we use root mean squared error (RMSE) as a performance metric for regression
\begin{align}
\text{RMSE}=\sqrt{\frac{1}{N_\text{test}N_L}\sum_{n\in \cI_\text{test}}\sum_{l=1}^{N_L}\left\|\bd_{n,l}-\hat{\bd}_{n,l}\right\|^2}.
\end{align}
In the experimental results, we compare following algorithms:
\begin{itemize}
 	\item \textbf{Pre:} pre-train the model in the training phase then re-train in the adaptation phase without CSI preprocessing.
	\item \textbf{TL:} transfer-learning-based model for people counting \cite{Khan:2023} and localization \cite{Noelia:2021} without CSI preprocessing. The model is pre-trained in the training phase, then only the FC layers of the CNN module are fine-tuned in the adaptation phase.
 	\item \textbf{Meta:} proposed meta-learning-based model without CSI preprocessing.
\item \textbf{Pre-CSI:} pre-trained model with CSI preprocessing.
\item \textbf{TL-CSI:} transfer-learning-based model with CSI preprocessing.
\item \textbf{Meta-CSI:} proposed meta-learning-based model with CSI preprocessing.
\end{itemize}

\begin{table}[t]
	\renewcommand{\arraystretch}{1.2}
	\captionsetup{justification=centering}
	\captionsetup{labelsep=newline}
	\caption{System parameters}
	\centering
	\label{table2}
	\begin{tabular}{l  l}
		\toprule
		Parameter & Value \\
		\midrule
		WiFi NIC &  Atheros AR$9380$\\
		Carrier frequency & $ 2.437 $ GHz\\
		Sampling frequency  & {$ 1 $ kHz}\\
		Number of subcarriers & $ 52 $ \\
		Number of spatial links  & $ 9 $\\
		Number of epochs & $ 20 $ \\
		Batch size & 64 \\
		Inner-loop rate & $10^{-1}$ \\
		Outer-loop rate & $10^{-4}$ \\
		\bottomrule
	\end{tabular}
\end{table}

\begin{figure}[t]
	\centering
	\begin{subfigure}{0.32\textwidth}
		\includegraphics[width=\textwidth]{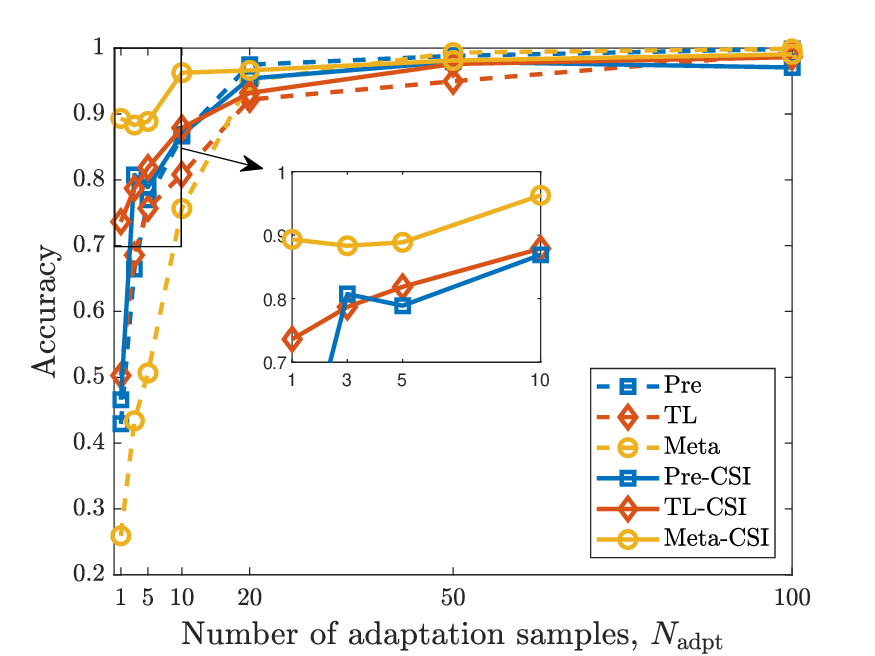}
		\caption{Small room.}
		\label{fig8:first}
	\end{subfigure}
	\hspace{1em}
	\begin{subfigure}{0.32\textwidth}
		\includegraphics[width=\textwidth]{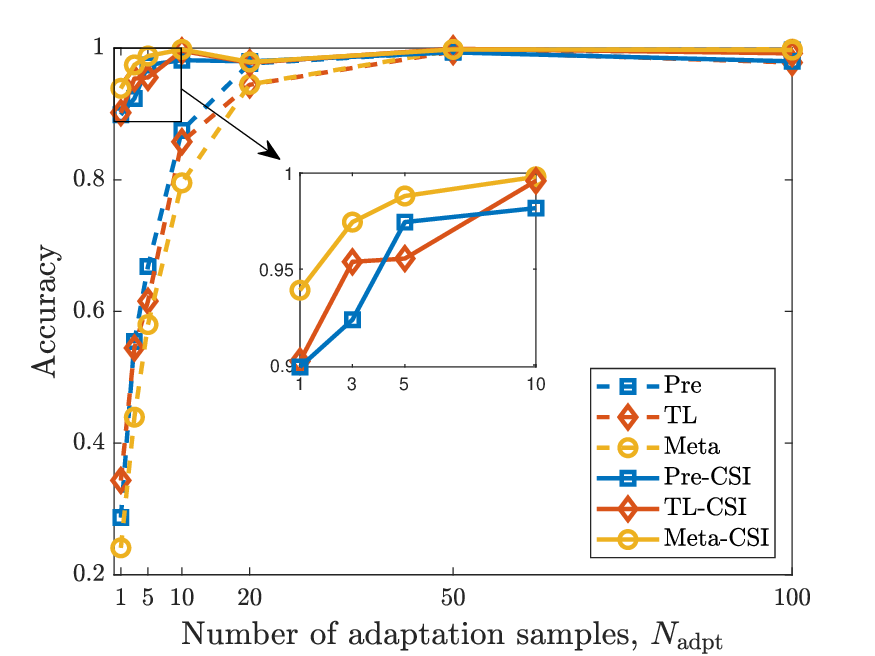}
		\caption{Large room.}
		\label{fig8:second}
	\end{subfigure}
	\hspace{1em}
	\begin{subfigure}{0.32\textwidth}
		\includegraphics[width=\textwidth]{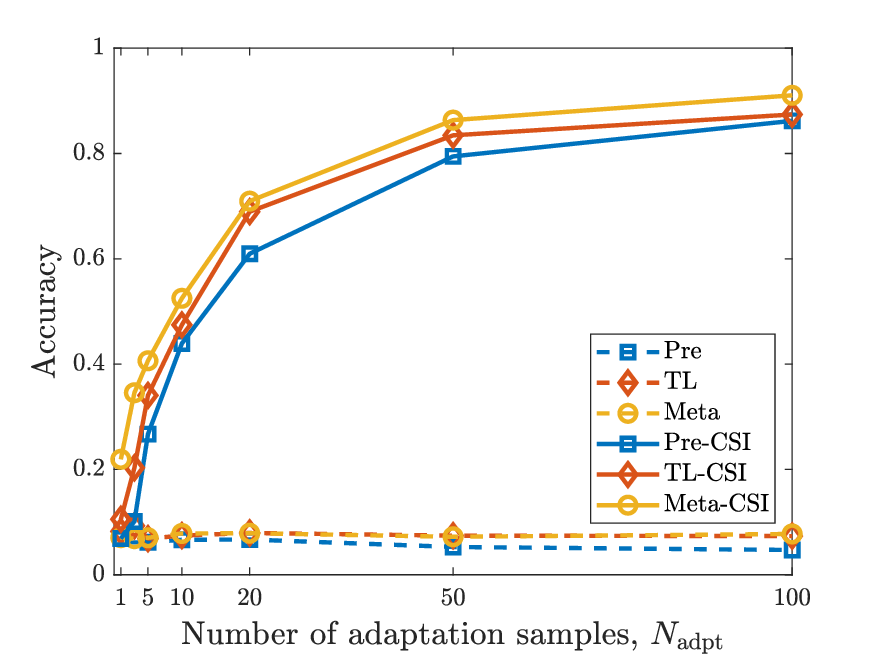}
		\caption{Open space.}
		\label{fig8:third}
	\end{subfigure}
	\caption{People counting accuracy vs. number of adaptation samples.}
	\label{fig8:figures}
\end{figure}

\begin{figure}[t]
	\centering
	\begin{subfigure}{0.3\textwidth}
		\includegraphics[width=\textwidth]{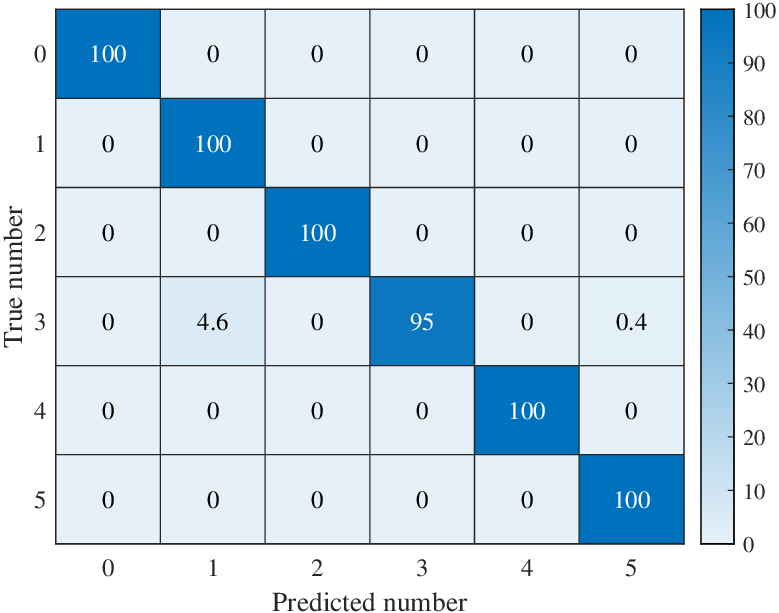}
		\caption{Small room.}
	\vspace{1em}
		\label{fig9:first}
	\end{subfigure}
    \hspace{1em}
	\begin{subfigure}{0.3\textwidth}
		\includegraphics[width=\textwidth]{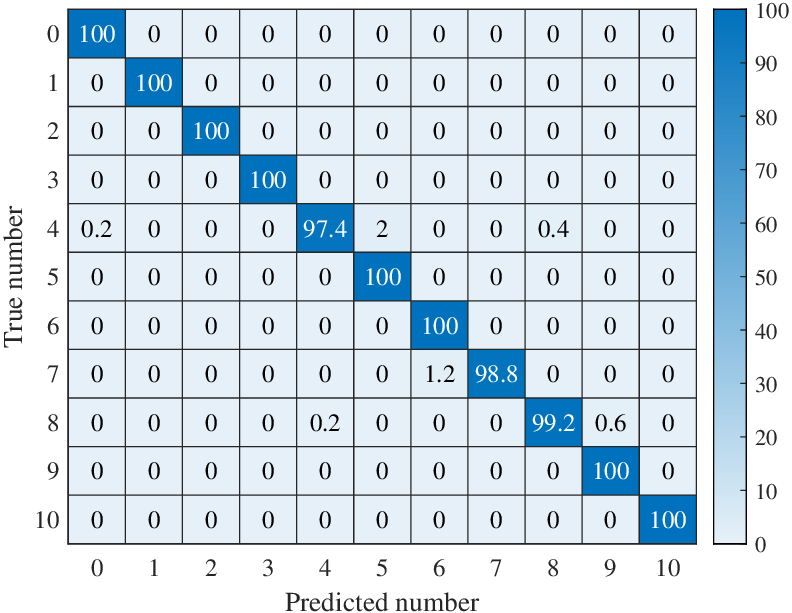}
		\caption{Large room.}
	\vspace{1em}
		\label{fig9:second}
	\end{subfigure}
    \hspace{1em}
	\begin{subfigure}{0.3\textwidth}
		\includegraphics[width=\textwidth]{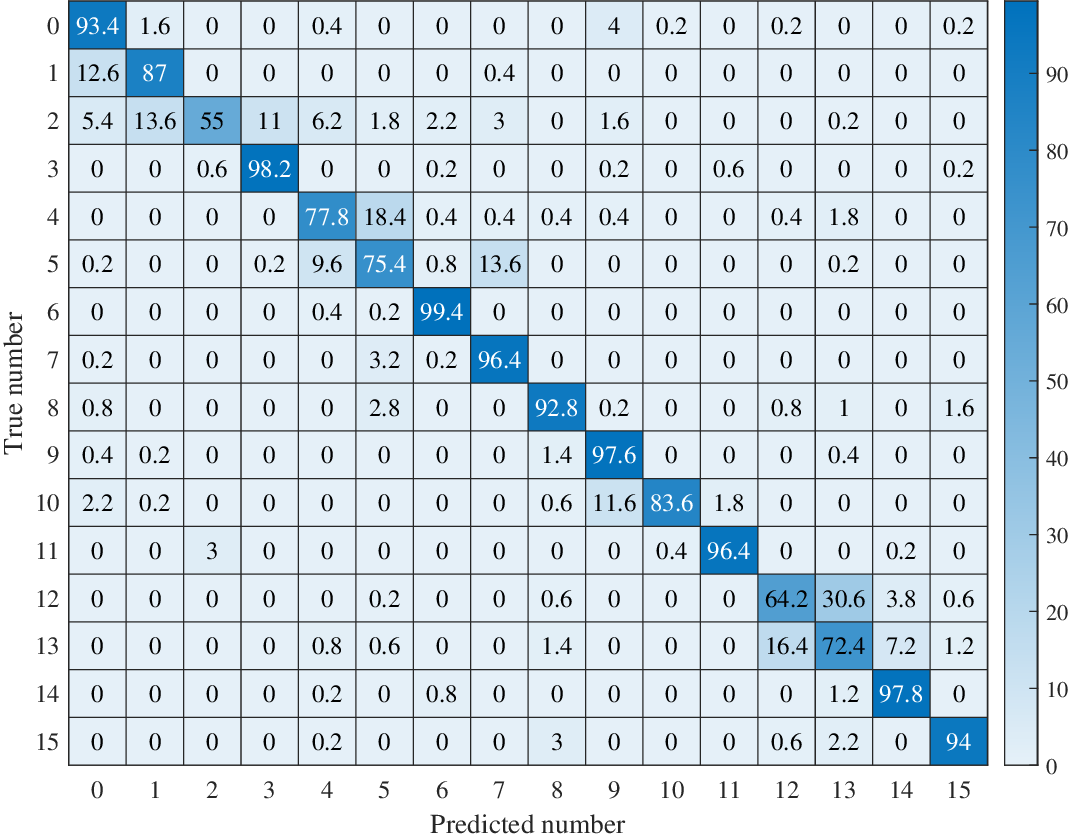}
		\caption{Open space.}
		\label{fig9:third}
	\end{subfigure}
	\caption{Confusion matrix of Meta-CSI for people counting when $N_\text{adpt}=50$.}
	\label{fig9:figures}
\end{figure}

\begin{figure}[t]
	\centering
	\begin{subfigure}[]{0.33\textwidth}
		\includegraphics[width=\textwidth]{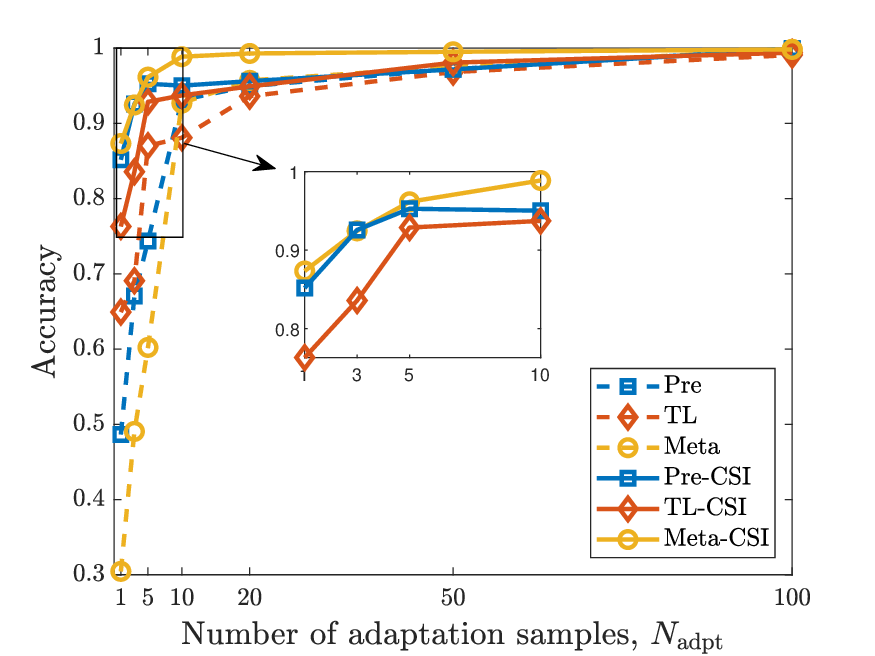}
		\caption{Small room.}
		\label{fig10:first}
	\end{subfigure}
	\begin{subfigure}[]{0.33\textwidth}
		\includegraphics[width=\textwidth]{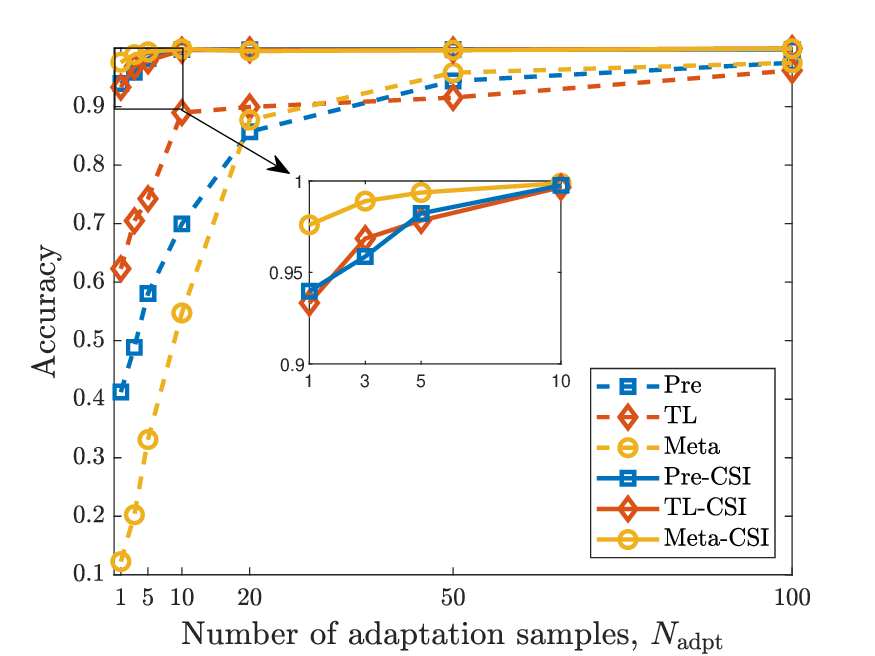}
		\caption{Large room.}
		\label{fig10:second}
	\end{subfigure}
    \begin{subfigure}[]{0.33\textwidth}
		\includegraphics[width=\textwidth]{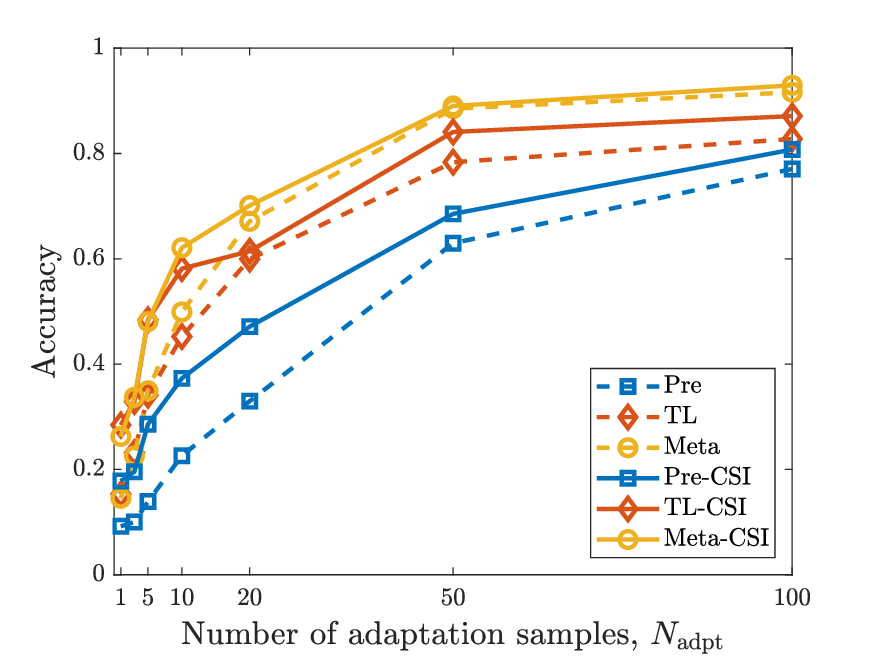}
		\caption{Open space.}
		\label{fig10:third}
	\end{subfigure}
	\caption{Localization accuracy vs. number of adaptation samples.}
	\label{fig10:figures}
\end{figure}

\begin{figure}[t]
	\centering
	\begin{subfigure}{0.30\textwidth}
		\includegraphics[width=\textwidth]{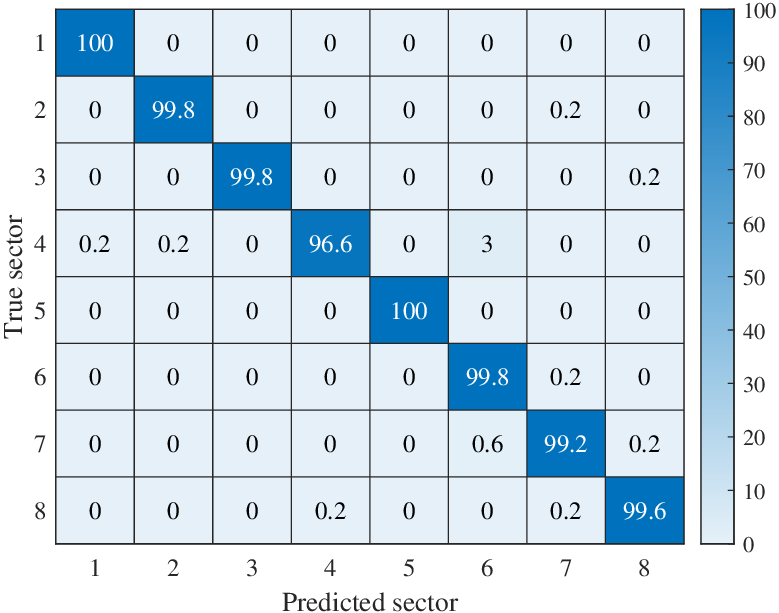}
		\caption{Small room.}
	\vspace{1em}
		\label{fig11:first}
	\end{subfigure}
	\hspace{1em}
	\begin{subfigure}{0.30\textwidth}
		\includegraphics[width=\textwidth]{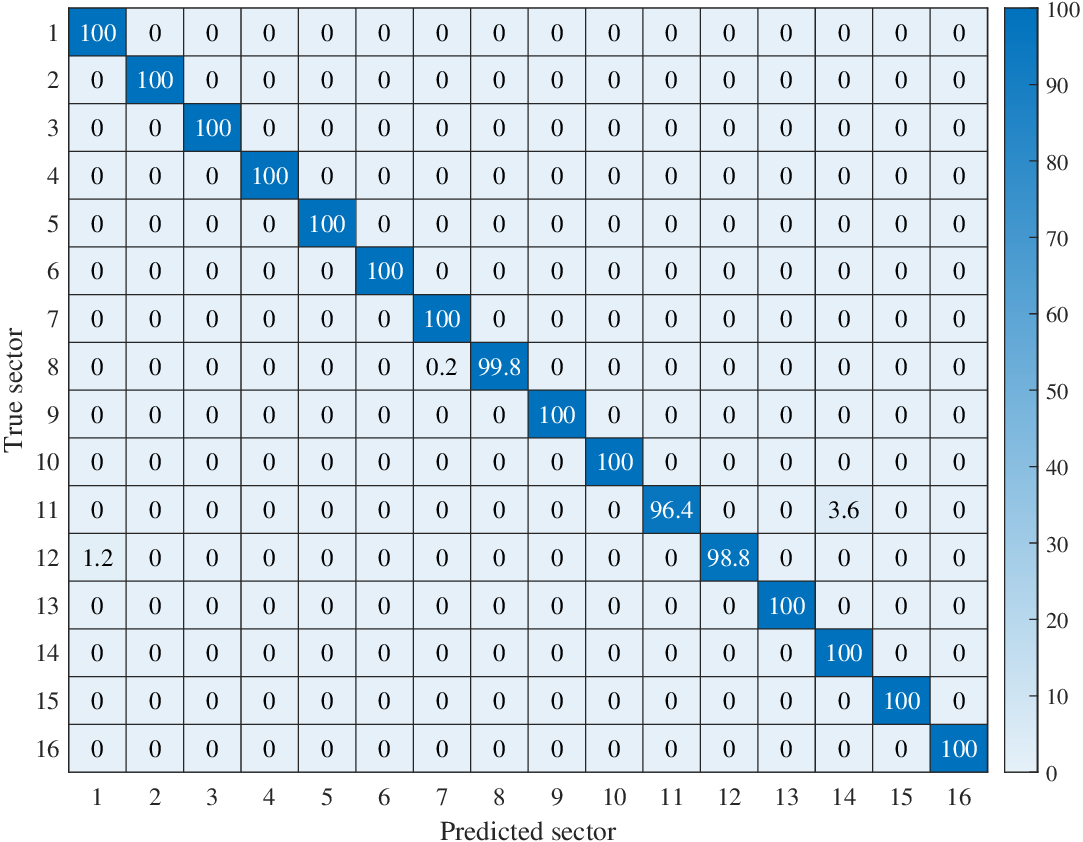}
		\caption{Large room.}
	\vspace{1em}
		\label{fig11:second}
	\end{subfigure}
    \hspace{1em}
    \begin{subfigure}{0.30\textwidth}
		\includegraphics[width=\textwidth]{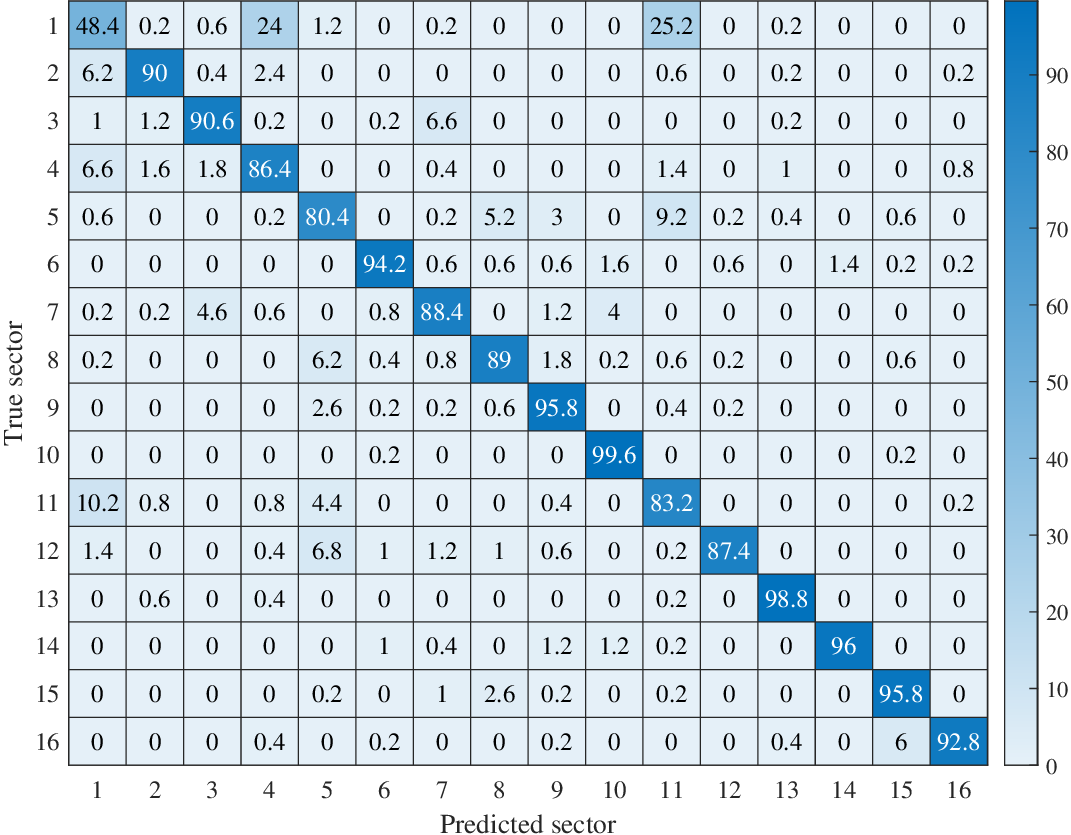}
		\caption{Open space.}
		\label{fig11:third}
	\end{subfigure}
	\caption{Confusion matrix of Meta-CSI for localization when $N_\text{adpt}=50$.}
	\label{fig11:figures}
\end{figure}

To evaluate the people counting model, we use 1000 training samples and 500 test samples for each class. We exploit 10000 training samples and 5000 test samples for the localization model. Also, we use the same number of adaptation samples $N_\textrm{adpt}$ for the pre-trained model, transfer-learning-based model, and meta-learning-based model for fair comparison.

\subsubsection{People Counting Model}
We consider the people counting problem with $C=6,11,$ and $16$ for the small room, large room, and open space, respectively. We set total $C-1$ positions for each environment, and locate an arbitrary number of people among the positions uniformly. In Fig. \ref{fig8:figures}, we plot the accuracy of pre-trained, transfer-learning-based, and meta-learning-based people counting models according to the number of adaptation samples. The models without CSI preprocessing show degraded performance with a small number of adaptation samples. Especially, the accuracy for the open space does not increase with $N_\text{adpt}$ at all. Longer and fewer propagation paths in such a wider environment make offsets more dominant. However, after CSI preprocessing, Meta-CSI clearly outperforms the other models for all environments. For the small and large rooms, we obtain higher than 95$\%$ accuracy of the people counting model with 20 adaptation samples in Meta-CSI. Even in the open space setting, Meta-CSI surpasses 80$\%$ accuracy with 50 adaptation samples. Thus, Meta-CSI can give accurate performance with a small number of adaptation samples.

\begin{figure}[tbp]
\centering
\begin{subfigure}{0.35\textwidth}
	\includegraphics[width=\textwidth]{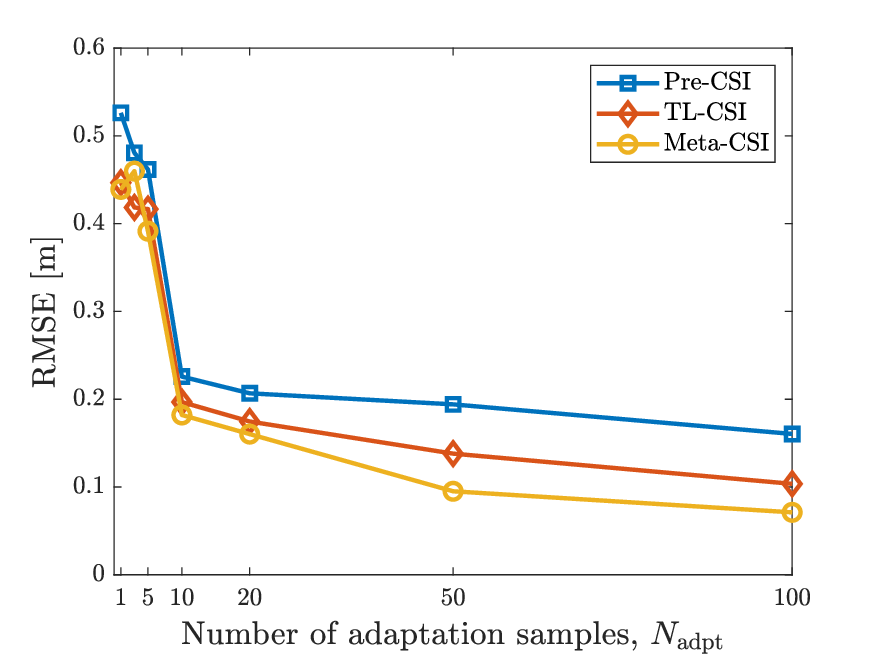}
	\caption{Same location.}
	\label{fig12:first}
\end{subfigure}
\hspace{1em}
\begin{subfigure}{0.35\textwidth}
	\includegraphics[width=\textwidth]{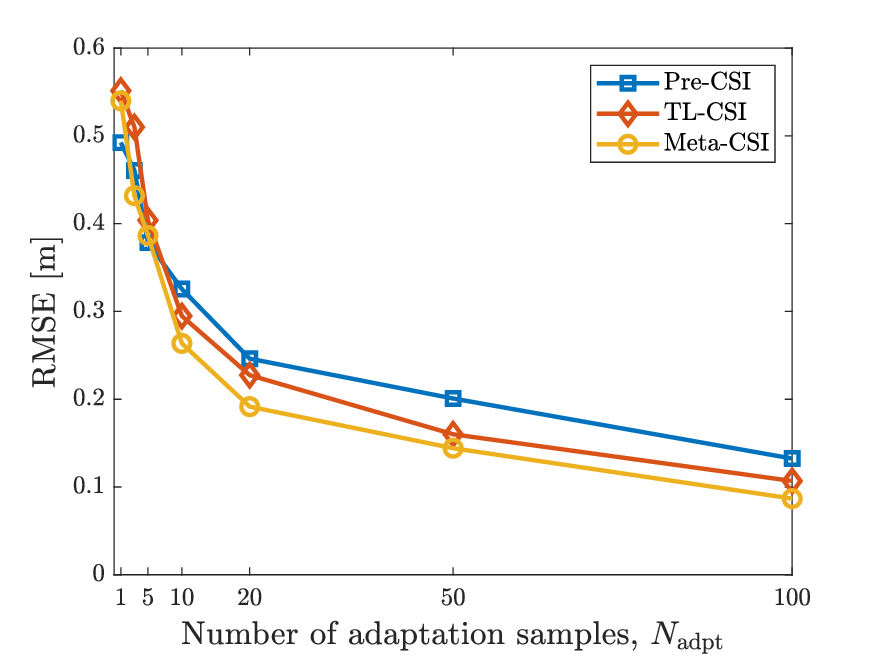}
	\caption{Different location.}
	\label{fig12:second}
\end{subfigure}
\caption{RMSE in large room vs. number of adaptation samples.}
\label{fig12:figures}
\end{figure}

\begin{figure}[tbp]
\centering
\begin{subfigure}{0.35\textwidth}
	\includegraphics[width=\textwidth]{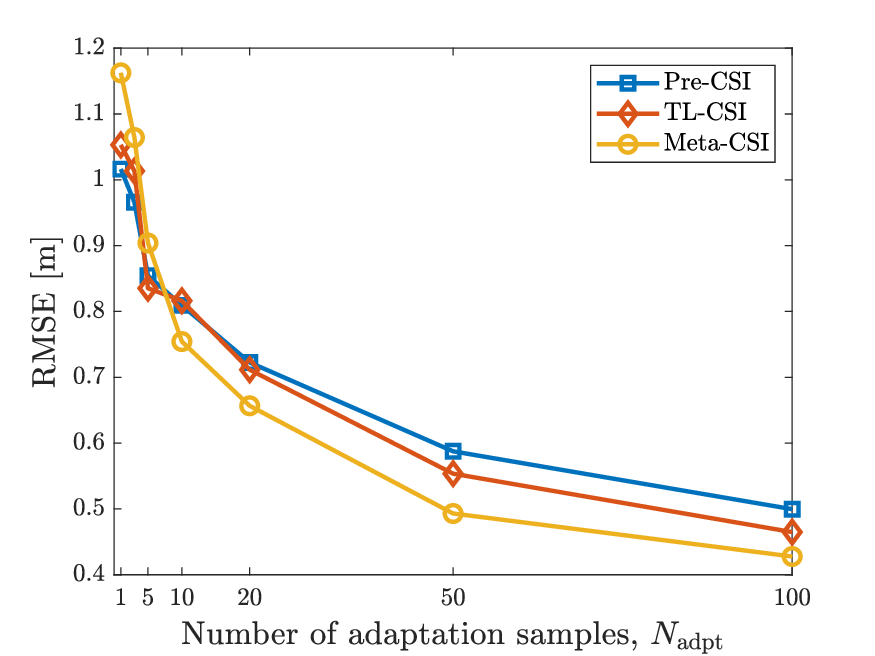}
	\caption{Same location.}
	\label{fig13:first}
\end{subfigure}
\hspace{1em}
\begin{subfigure}{0.35\textwidth}
	\includegraphics[width=\textwidth]{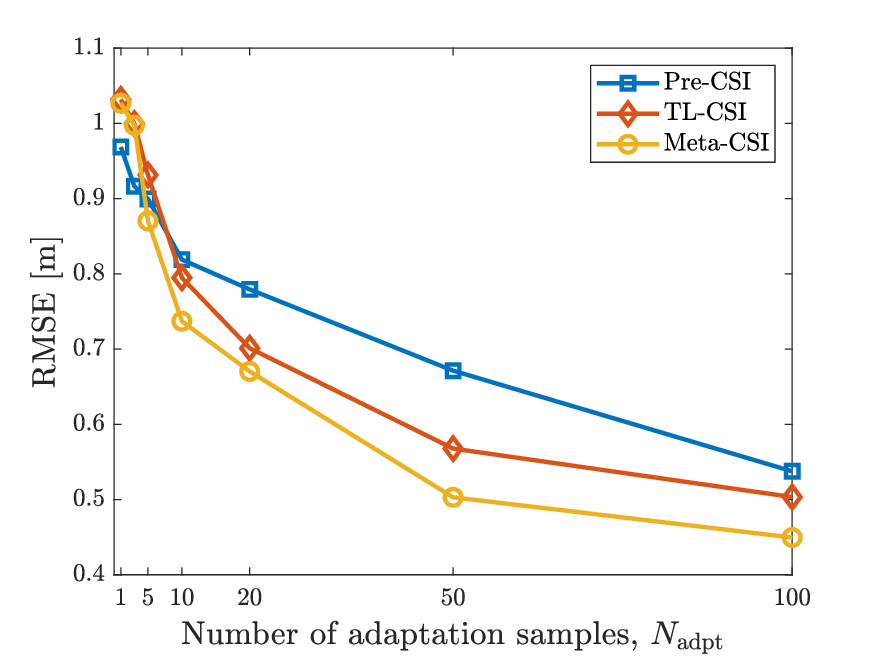}
	\caption{Different location.}
	\label{fig13:second}
\end{subfigure}
\caption{RMSE in open space vs. number of adaptation samples.}
\label{fig13:figures}
\end{figure}

We also plot the confusion matrix of Meta-CSI for people counting when $N_\text{adpt}=50$ in Fig. \ref{fig9:figures}. The confusion matrix is the classification table; each row of matrix represents the true values, while each column represents the predicted values. Each entry of the matrix expresses the prediction ratio of each class. For both the small room and large room, the accuracy of each class is nearly 100$\%$. Moreover, the accuracy is close to 90$\%$ in the open space.

\subsubsection{Localization Model (Classification)}\label{localization model classification} 
We also consider the localization problem for single person as classification. The small room is equally divided into $C=4\times2=8$ sectors, and the large room and open space are divided into $C=4\times4=16$ sectors. We assume that only one person can stand at the center of each sector, and the size of sectors is set to $1.2$ m $\times$ $1.2$ m. Similar to the people counting model, we exploit the meta-learning algorithm to classify each sector. In Fig. \ref{fig10:figures}, we show the accuracy of localization models according to the number of adaptation samples. Although Meta and the baselines without CSI preprocessing show inconsistent accuracy for different environments, Meta-CSI steadily has the best performance. Fig. \ref{fig11:figures} depicts the confusion matrix of Meta-CSI for localization when $N_\text{adpt}=50$. The confusion matrices show that the accuracy of each class is almost 100$\%$ in both the small room and large room, while it is about 90$\%$ in the open space. For incorrect prediction, however, the minimum localization error cannot be smaller than a sector size. For example, the error for large room is larger than $1.2$ m since we set each sector size as $1.2$ m $\times$ $1.2$ m. Therefore, we employ the regression model for more accurate localization in Section~\ref{localization model regression}.


\subsubsection{Localization Model (Regression)} \label{localization model regression}
Different from Section \ref{localization model classification}, we consider a two-person localization problem in the large room and open space, and the output of learning models is the location of two people, i.e., $N_L=2$. We set two different scenarios, i.e., same location and different location. The same location implies the locations of people of the training dataset and the test dataset are the same. The different location implies a more practical case that the locations of people of the training dataset and the test dataset~are~different. Specifically, we collect the training samples where the two people are located among the designated positions with uniform spacing of $1.2$ m, and the test datasets for the different location do not include the positioning cases in the training dataset.

Fig. \ref{fig12:figures} shows the RMSEs of localization models with CSI preprocessing according to the number of adaptation samples in the large room. Meta-CSI outperforms Pre-CSI and TL-CSI, and the RMSE value of Meta-CSI is less than 0.2 m with $N_\text{adpt}\geq 20$. Also, even in the different location scenario, the proposed Meta-CSI has moderate performance with a small number of adaptation~samples.

In contrast to the large room results, Fig. \ref{fig13:figures} indicates that the RMSE values in the open space scenario are generally higher, likely due to increased signal fluctuations and environmental interference. Despite these more challenging conditions, all three methods show a gradual decrease in RMSE as the number of adaptation samples increases. In particular, Meta-CSI retains the lowest RMSE curve throughout the adaptation process, which demonstrates its robustness in more demanding environments. 

\begin{figure}[tbp]
\centering
\begin{subfigure}{0.35\textwidth}
	\includegraphics[width=\textwidth]{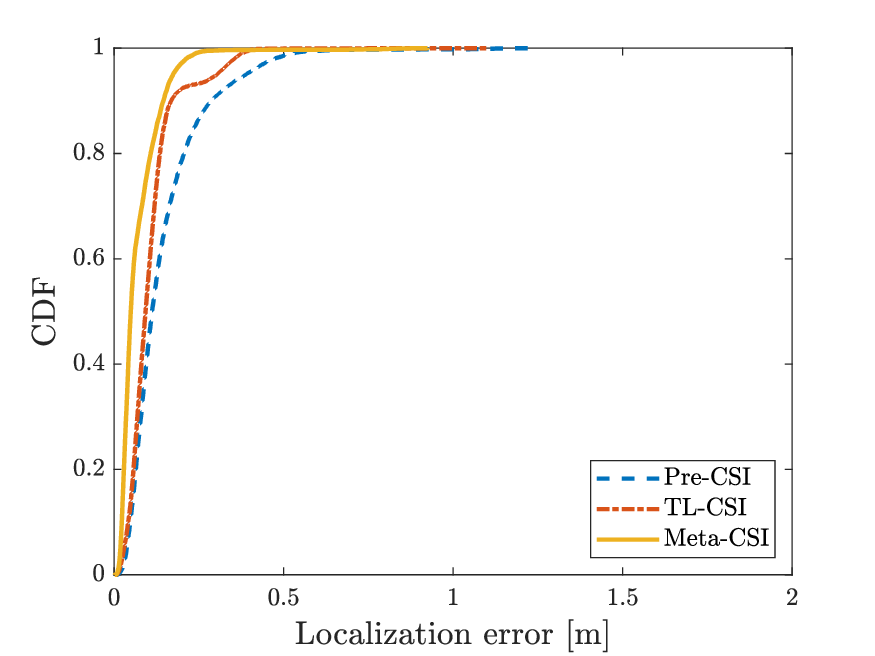}
	\caption{Same location.}
	\label{fig14:first}
\end{subfigure}
\hspace{1em}
\begin{subfigure}{0.35\textwidth}
	\includegraphics[width=\textwidth]{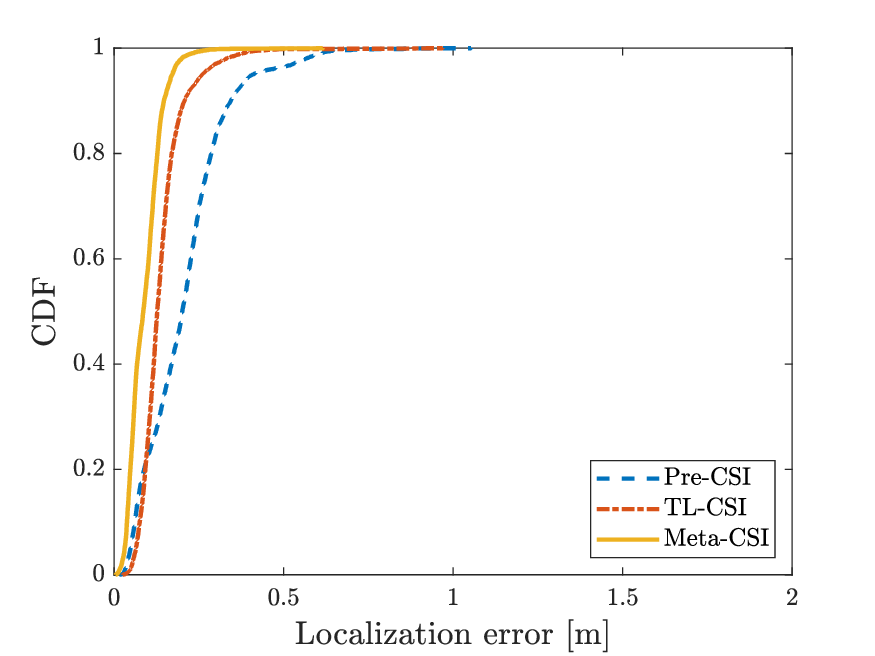}
	\caption{Different location.}
	\label{fig14:second}
\end{subfigure}
\caption{CDF of localization error in large room with $N_\textrm{adpt}=50$.}
\label{fig14:figures}
\end{figure}

Fig. \ref{fig14:figures} reveals the cumulative distribution functions (CDFs) of localization error with the number of adaptation samples $N_\textrm{adpt}=50$. 
Note that the localization error is defined as 
\begin{align}
\text{Localization error}=\frac{1}{N_L}\sum_{l=1}^{N_L}\left\|\bd_{n,l}-\hat{\bd}_{n,l}\right\|.
\end{align}The figure clearly shows that Meta-CSI has superior performance compared to the other models. In the same location, $90 \%$ CDF levels of Meta-CSI, TL-CSI, and Pre-CSI  are $0.1452$ m, $0.1684$ m, and $0.2871$ m, respectively. In the different location, $90\%$ CDF levels of Meta-CSI, TL-CSI, and Pre-CSI are $0.1469$ m, $0.2077$ m, and $0.3443$ m. Although the localization errors in the different location are slightly higher than those in the same location, Meta-CSI performs well in the different location. This result shows that the proposed Meta-CSI can be used in practical scenarios.

\begin{figure}[tbp]
\centering
\begin{subfigure}{0.35\textwidth}
	\includegraphics[width=\textwidth]{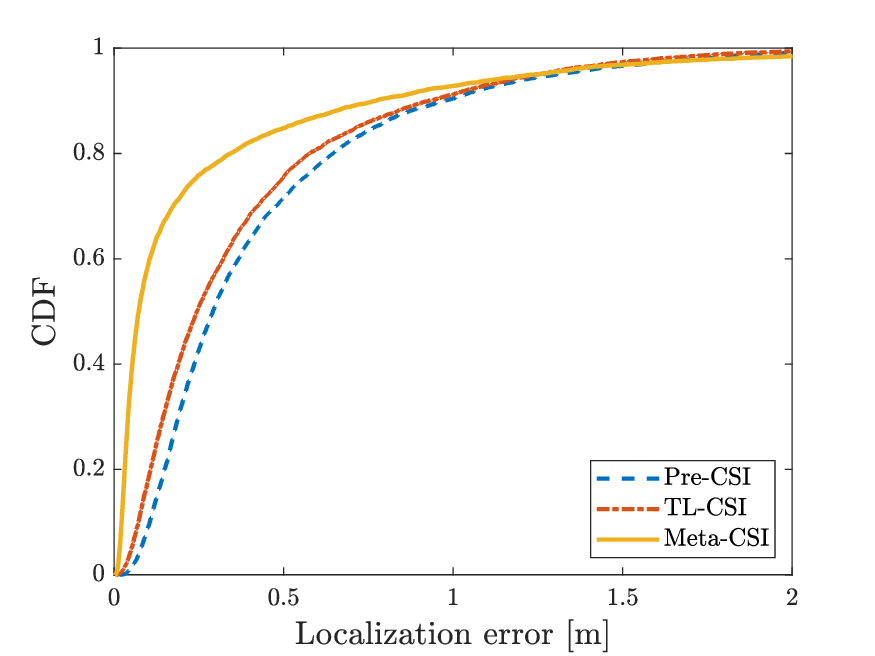}
	\caption{Same location.}
	\label{fig15:first}
\end{subfigure}
\hspace{1em}
\begin{subfigure}{0.35\textwidth}
	\includegraphics[width=\textwidth]{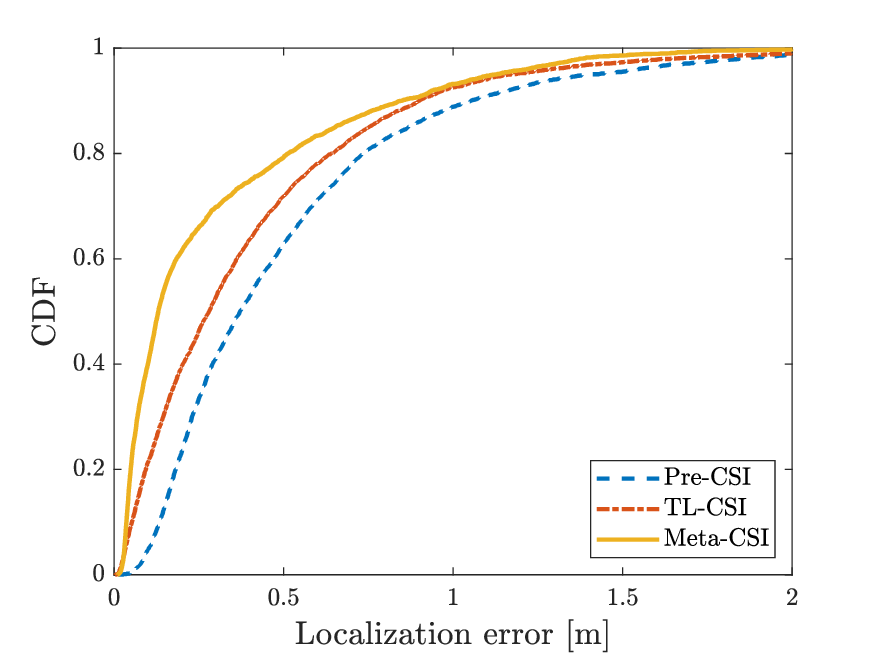}
	\caption{Different location.}
	\label{fig15:second}
\end{subfigure}
\caption{CDF of localization error in open space with $N_\textrm{adpt}=50$.}
\label{fig15:figures}
\end{figure}

As shown in Fig. \ref{fig15:figures}, the open space environment yields larger overall localization errors compared to the large room, even with 50 adaptation samples. At the 90$\%$ CDF level, the localization errors for Meta-CSI, TL-CSI, and Pre-CSI are 0.7701 m, 0.9267 m, and 0.9737 m, respectively, in the same location scenario, and 0.8442 m, 0.8993 m, and 1.052 m, respectively, in the different location scenario. Despite the challenging nature of the open space environment, Meta-CSI still gives the lowest overall error distribution. These findings highlight the importance of robust adaptation techniques in open spaces, where multipath effects can substantially impact localization accuracy.



\section{Conclusion and Discussion}\label{sec6}
In this paper, we configured the people counting and localization systems employing CSI from the commodity WiFi NICs. The preprocessing of CSI was proposed to remove the amplitude and phase offsets from erroneous CSI. The proposed meta-learning-based model optimizes its parameters into general ones for human sensing, which makes the model adaptive to different measurement environments. The experimental results revealed that the proposed meta-learning-based model with CSI preprocessing has notable gain over the benchmark models with a small amount of adaptation samples in both people counting and localization problems.

While the proposed scheme demonstrates its ability for adaptive people counting and localization in wide spaces without relying on extensive datasets, the limitation and challenges remain. Labeling training data to distinguish various cases involving people and locations is a manual and exhaustive task. Additionally, nonlinear offset components in the measured CSI, which vary across time-frequency resources, still exist. To enhance scalability, the proposed meta-learning-based model requires improvement with a larger scale of data collection in different environments and with high computing resources for fast training and adaptation processes. As a possible future work, we can adopt unsupervised learning for the proposed model without data labeling, and CSI preprocessing for offset removal can also be improved by treating its nonlinear components.



\ifCLASSOPTIONcaptionsoff
  \newpage
\fi



\bibliographystyle{IEEEtran}
\bibliography{refs_all}

\begin{thebibliography}{10}
\providecommand{\url}[1]{#1}
\csname url@samestyle\endcsname
\providecommand{\newblock}{\relax}
\providecommand{\bibinfo}[2]{#2}
\providecommand{\BIBentrySTDinterwordspacing}{\spaceskip=0pt\relax}
\providecommand{\BIBentryALTinterwordstretchfactor}{4}
\providecommand{\BIBentryALTinterwordspacing}{\spaceskip=\fontdimen2\font plus
\BIBentryALTinterwordstretchfactor\fontdimen3\font minus
  \fontdimen4\font\relax}
\providecommand{\BIBforeignlanguage}[2]{{%
\expandafter\ifx\csname l@#1\endcsname\relax
\typeout{** WARNING: IEEEtran.bst: No hyphenation pattern has been}%
\typeout{** loaded for the language `#1'. Using the pattern for}%
\typeout{** the default language instead.}%
\else
\language=\csname l@#1\endcsname
\fi
#2}}
\providecommand{\BIBdecl}{\relax}
\BIBdecl

\bibitem{Teixeira:2010}
T.~{Teixeira}, G.~{Dublon}, and A.~{Savvides}, ``{A Survey of Human-Sensing:
  Methods for Detecting Presence, Count, Location, Track, and Identity},''
  \emph{ACM Computing Surveys}, vol.~5, no.~1, pp. 59--69, 2010.

\bibitem{Liu:2019}
Z.~{Liu}, X.~{Liu}, J.~{Zhang}, and K.~{Li}, ``{Opportunities and Challenges of
  Wireless Human Sensing for the Smart IoT World: A Survey},'' \emph{IEEE
  Network}, vol.~33, no.~5, pp. 104--110, 2019.

\bibitem{Liu:2020a}
J.~{Liu}, H.~{Liu}, Y.~{Chen}, Y.~{Wang}, and C.~{Wang}, ``{Wireless Sensing
  for Human Activity: A Survey},'' \emph{IEEE Communications Surveys \&
  Tutorials}, vol.~22, no.~3, pp. 1629--1645, Oct. 2020.

\bibitem{Liu:2020b}
J.~{Liu}, G.~{Teng}, and F.~{Hong}, ``{Human Activity Sensing with Wireless
  Signals: A Survey},'' \emph{Sensors}, vol.~20, no.~4, Feb. 2020.

\bibitem{Yang:2003}
D.~B. {Yang}, H.~H. {González-Baños}, and J.~G. {Guibas}, ``{Counting People
  in Crowds with a Real-Time Network of Simple Image Sensors},''
  \emph{Proceedings Ninth IEEE International Conference on Computer Vision},
  pp. 122--129, Oct. 2003.

\bibitem{Aggarwal:2011}
J.~{Aggarwal} and M.~S. {Ryoo}, ``{Human Activity Analysis: A Review},''
  \emph{ACM Computing Surveys}, vol.~43, no.~3, Apr. 2011.

\bibitem{Ke:2013}
S.~R. {Ke}, H.~L.~U. {Thuc}, Y.~J. {Lee}, J.~N. {Hwang}, J.~H. {Yoo}, and K.~H.
  {Choi}, ``{A Review on Video-Based Human Activity Recognition},''
  \emph{Computers}, vol.~2, no.~2, pp. 88--131, Jun. 2013.

\bibitem{Adib:2015}
F.~{Adib}, Z.~{Kabelac}, and D.~{Katabi}, ``{Multi-Person Localization via RF
  Body Reflections},'' \emph{12th USENIX Symposium on Networked Systems Design
  and Implementation (NSDI 15)}, pp. 279--292, May 2015.

\bibitem{Lien:2016}
J.~{Lien}, N.~{Gillian}, M.~E. {Karagozler}, P.~{Amihood}, C.~{Schwesig},
  E.~{Olson}, H.~{Raja}, and I.~{Poupyrev}, ``{Soli: Ubiquitous Gesture Sensing
  with Millimeter Wave Radar},'' \emph{ACM Transactions on Graphics}, vol.~35,
  no.~4, Jul. 2016.

\bibitem{Alizadeh:2019}
M.~{Alizadeh}, G.~{Shaker}, J.~C.~M. {De Almeida}, P.~P. {Morita}, and
  S.~{Safavi-Naeini}, ``{Remote Monitoring of Human Vital Signs Using mm-Wave
  FMCW Radar},'' \emph{IEEE Access}, vol.~7, pp. 54\,958--54\,968, Apr. 2019.

\bibitem{Khan:2023}
D.~Khan and I.~W.-H. Ho, ``{CrossCount: Efficient Device-Free Crowd Counting by
  Leveraging Transfer Learning},'' \emph{IEEE Internet of Things Journal},
  vol.~10, no.~5, pp. 4049--4058, Mar. 2023.

\bibitem{Soltanaghaei:2020}
E.~{Soltanaghaei}, R.~A. {Sharma}, Z.~{Wang}, A.~{Chittilappilly}, A.~{Luong},
  E.~{Giler}, K.~{Hall}, S.~{Elias}, and A.~{Rowe}, ``{Robust and Practical
  WiFi Human Sensing Using On-Device Learning with a Domain Adaptive Model},''
  \emph{Proceedings of the 7th ACM International Conference on Systems for
  Energy-Efficient Buildings, Cities, and Transportation}, pp. 150--159, Nov.
  2020.

\bibitem{Zhao:2019}
Y.~{Zhao}, S.~{Liu}, F.~{Xue}, B.~{Chen}, and X.~{Chen}, ``{DeepCount: Crowd
  Counting with Wi-Fi Using Deep Learning},'' \emph{Journal of Communications
  and Information Networks}, vol.~4, no.~3, pp. 38--52, Sep. 2019.

\bibitem{Wang:2017a}
W.~{Wang}, A.~X. {Liu}, M.~{Shahzad}, K.~{Ling}, and S.~{Lu}, ``{Device-Free
  Human Activity Recognition Using Commercial WiFi Devices},'' \emph{IEEE
  Journal on Selected Areas in Communications}, vol.~35, no.~5, pp. 1118--1131,
  May 2017.

\bibitem{Jiang:2020}
W.~{Jiang}, H.~{Xue}, C.~{Miao}, S.~{Wang}, S.~{Lin}, C.~{Tian}, S.~{Murali},
  H.~{Hu}, and L.~{Su}, ``{Towards 3D Human Pose Construction Using WiFi},''
  \emph{Proceedings of the 26th Annual International Conference on Mobile
  Computing and Networking}, pp. 1--14, Sep. 2020.

\bibitem{Yousefi:2017}
S.~{Yousefi}, H.~{Narui}, S.~{Dayal}, S.~{Ermon}, and S.~{Valaee}, ``{A Survey
  on Behavior Recognition Using WiFi Channel State Information},'' \emph{IEEE
  Communications Magazine}, vol.~55, no.~10, pp. 98--104, Oct. 2017.

\bibitem{Bu:2020}
Q.~{Bu}, G.~{Yang}, X.~{Ming}, T.~{Zhang}, J.~{Feng}, and J.~{Zhang}, ``{Deep
  Transfer Learning for Gesture Recognition with WiFi Signals},''
  \emph{Personal and Ubiquitous Computing}, pp. 1--12, Jan. 2020.

\bibitem{Niu:2021}
K.~{Niu}, F.~{Zhang}, X.~{Wang}, Q.~{Lv}, H.~{Luo}, and D.~{Zhang},
  ``{Understanding WiFi Signal Frequency Features for Position-Independent
  Gesture Sensing},'' \emph{IEEE Transactions on Mobile Computing}, Mar. 2021.

\bibitem{Arshad:2019}
S.~{Arshad}, C.~{Feng}, R.~{Yu}, and Y.~{Liu}, ``{Leveraging Transfer Learning
  in Multiple Human Activity Recognition Using WiFi Signal},'' \emph{2019 IEEE
  20th International Symposium on "A World of Wireless, Mobile and Multimedia
  Networks" (WoWMoM)}, pp. 1--10, Jun. 2019.

\bibitem{Noelia:2021}
N.~Hernández, I.~Parra, H.~Corrales, R.~Izquierdo, A.~L. Ballardini,
  C.~Salinas, and I.~García, ``{WiFiNet: WiFi-Based Indoor Localisation Using
  CNNs},'' \emph{Expert Systems with Applications}, vol. 177, p. 114906, 2021.

\bibitem{Foliadis:2021}
A.~{Foliadis}, M.~H.~C. {Garcia}, R.~A. {Stirling-Gallacher}, and R.~S.
  {Thomä}, ``{Reliable Deep Learning Based Localization with CSI Fingerprints
  and Multiple Base Stations},'' \emph{arXiv:2111.11839}, 2021.

\bibitem{Qian:2017}
K.~{Qian}, C.~{Wu}, Z.~{Yang}, Y.~{Liu}, and K.~{Jamieson}, ``{Widar:
  Decimeter-Level Passive Tracking via Velocity Monitoring with Commodity
  Wi-Fi},'' \emph{Proceedings of the 18th ACM International Symposium on Mobile
  Ad Hoc Networking and Computing}, pp. 1--10, Jul. 2017.

\bibitem{Qian:2018}
K.~{Qian}, C.~{Wu}, Y.~{Zhang}, G.~{Zhang}, Z.~{Yang}, and Y.~{Liu},
  ``{Widar2.0: Passive Human Tracking with a Single Wi-Fi Link},''
  \emph{Proceedings of the 16th Annual International Conference on Mobile
  Systems, Applications, and Services}, pp. 350--361, Jun. 2018.

\bibitem{Tan:2019}
S.~{Tan}, L.~{Zhang}, Z.~{Wang}, and J.~{Yang}, ``{MultiTrack: Multi-User
  Tracking and Activity Recognition Using Commodity WiFi},'' \emph{Proceedings
  of the 2019 CHI Conference on Human Factors in Computing Systems}, pp. 1--12,
  May 2019.

\bibitem{Choi:2021}
H.~{Choi}, T.~{Matsui}, S.~{Misaki}, A.~{Miyaji}, M.~{Fujimoto}, and
  K.~{Yasumoto}, ``{Simultaneous Crowd Estimation in Counting and Localization
  Using WiFi CSI},'' \emph{2021 International Conference on Indoor Positioning
  and Indoor Navigation (IPIN)}, pp. 1--8, 2021.

\bibitem{Yan:2021}
J.~{Yan}, L.~{Wan}, W.~{Wei}, X.~{Wu}, W.~P. {Zhu}, and D.~P.~K. {Lun},
  ``{Device-Free Activity Detection and Wireless Localization Based on CNN
  Using Channel State Information Measurement},'' \emph{IEEE Sensors Journal},
  vol.~21, no.~21, pp. 24\,482--24\,494, Nov. 2021.

\bibitem{Xie:2019}
Y.~{Xie}, Z.~{Li}, and M.~{Li}, ``{Precise Power Delay Profiling with Commodity
  Wi-Fi},'' \emph{IEEE Transactions on Mobile Computing}, vol.~18, no.~6, pp.
  1342--1355, Jun. 2019.

\bibitem{Wang:2017c}
X.~{Wang}, C.~{Yang}, and S.~{Mao}, ``Tensorbeat: Tensor decomposition for
  monitoring multiperson breathing beats with commodity wifi,'' \emph{ACM
  Trans. Intell. Syst. Technol.}, vol.~9, no.~1, Sep. 2017.

\bibitem{Wang:2017b}
X.~{Wang}, L.~{Gao}, S.~{Mao}, and S.~{Pandey}, ``{CSI-Based Fingerprinting for
  Indoor Localization: A Deep Learning Approach},'' \emph{IEEE Transactions on
  Vehicular Technology}, vol.~66, no.~1, pp. 763--776, Jan. 2017.

\bibitem{Finn:2017}
C.~{Finn}, P.~{Abbeel}, and S.~{Levine}, ``{Model-Agnostic Meta-Learning for
  Fast Adaptation of Deep Networks},'' \emph{Proceedings of the 34th
  International Conference on Machine Learning}, vol.~70, pp. 1126--1135, Aug.
  2017.

\bibitem{Speth:1999}
M.~{Speth}, S.~{Fechtel}, G.~{Fock}, and H.~{Meyr}, ``Optimum receiver design
  for wireless broad-band systems using ofdm. i,'' \emph{IEEE Transactions on
  Communications}, vol.~47, no.~11, pp. 1668--1677, Nov. 1999.

\bibitem{Pan:2010}
S.~J. {Pan} and Q.~{Yang}, ``{A Survey on Transfer Learning},'' \emph{IEEE
  Transactions on Knowledge and Data Engineering}, vol.~22, no.~10, pp.
  1345--1359, Oct. 2010.

\bibitem{Zheng:2017}
Y.~{Zheng}, C.~{Wu}, K.~{Qian}, Z.~{Yang}, and Y.~{Liu}, ``{Detecting Radio
  Frequency Interference for CSI Measurements on COTS WiFi Devices},'' in
  \emph{2017 IEEE International Conference on Communications (ICC)}, 2017, pp.
  1--6.

\bibitem{Huang:2020}
J.~{Huang}, B.~{Liu}, C.~{Chen}, H.~{Jin}, Z.~{Liu}, C.~{Zhang}, and N.~{Yu},
  ``{Towards Anti-Interference Human Activity Recognition Based on WiFi
  Subcarrier Correlation Selection},'' \emph{IEEE Transactions on Vehicular
  Technology}, vol.~69, no.~6, pp. 6739--6754, Jun. 2020.

\bibitem{Ahmet:2020}
A.~M. Elbir, A.~Papazafeiropoulos, P.~Kourtessis, and S.~Chatzinotas, ``{Deep
  Channel Learning for Large Intelligent Surfaces Aided mm-Wave Massive MIMO
  Systems},'' \emph{IEEE Wireless Communications Letters}, vol.~9, no.~9, pp.
  1447--1451, 2020.

\bibitem{Hunger:2005}
R.~Hunger, \emph{Floating point operations in matrix-vector calculus}.\hskip
  1em plus 0.5em minus 0.4em\relax Munich University of Technology, Inst. for
  Circuit Theory and Signal, 2005.

\bibitem{Mizutani:2001}
E.~{Mizutani} and S.~E. {Dreyfus}, ``On complexity analysis of supervised
  {MLP}-learning for algorithmic comparisons,'' in \emph{IJCNN'01.
  International Joint Conference on Neural Networks. Proceedings (Cat.
  No.01CH37222)}, vol.~1, Jul. 2001, pp. 347--352.

\bibitem{Taghavi:2019}
M.~Taghavi and M.~Shoaran, ``Hardware complexity analysis of deep neural
  networks and decision tree ensembles for real-time neural data
  classification,'' in \emph{2019 9th International IEEE/EMBS Conference on
  Neural Engineering (NER)}, Mar. 2019, pp. 407--410.

\bibitem{Kim:2021}
H.~Kim, S.~Kim, H.~Lee, C.~Jang, Y.~Choi, and J.~Choi, ``Massive {MIMO} channel
  prediction: {Kalman} filtering vs. machine learning,'' \emph{IEEE
  Transactions on Communications}, vol.~69, no.~1, pp. 518--528, Jan. 2021.

\bibitem{Kim:2023}
H.~Kim, J.~Choi, and D.~J. Love, ``Massive {MIMO} channel prediction via
  meta-learning and deep denoising: Is a small dataset enough?'' \emph{IEEE
  Transactions on Wireless Communications}, vol.~22, no.~12, pp. 9278--9290,
  Dec. 2023.

\end{thebibliography}

\end{document}